\documentclass[12pt]{article}
\usepackage[margin=2.5cm]{geometry}
\usepackage[utf8]{inputenc}
\usepackage{graphicx}
\usepackage{float}
\usepackage{amsmath, amssymb, amsfonts}
\usepackage{tikz}
\usepackage{subcaption}
\usepackage{caption}

\usepackage{algorithm,algpseudocode}

\usepackage{amsthm}
\usepackage{enumitem}
\usepackage{balance}
\usepackage{lineno}

\usepackage{setspace}
\begin{document}

\title{\textbf{Exploring Bounded Component Analysis Using an $\ell_\infty$ Norm Criterion}}

\author{
Renan D. B. Brotto\thanks{Corresponding author: rbrotto@gmail.com. This work is not associated with author’s professional activities at Samsung.} \\
\small School of Electrical and Computer Engineering, University of Campinas \\
\small Samsung R\&D Institute Brazil - SRBR \\
\and
Kenji Nose-Filho \\
\small Center of Engineering, Modeling and Applied Social Sciences, \\
\small Federal University of ABC (UFABC) \\
\and
João M. T. Romano \\
\small School of Electrical and Computer Engineering, University of Campinas
}

\date{} 

\maketitle

\begin{center}
This manuscript is made available under the Creative Commons Attribution-NonCommercial-NoDerivatives 4.0 International License (CC BY-NC-ND 4.0). The final published version is available at https://doi.org/10.1016/j.dsp.2024.104696
\end{center}

\newpage

\begin{abstract}
In this paper we propose a new criterion for the Blind Source Separation (BSS) of antisparse bounded sources, based on the sum of the $\ell_\infty$-norm of the sources. Based on the observation that the mixing process of bounded sources with any mixing matrix with unitary Frobenius norm will increase the $\ell_\infty$-norm of the sources, unless it is the identity matrix, the minimization of the sum of the $\ell_\infty$-norm of the sources can be used for the estimation of a separation matrix. To that, a Principle Component Analysis technique followed by a Givens Rotations based optimization method can be used for the separation of independent bounded sources. Also, the Givens Rotations based optimization method can be used for the separation of correlated bounded sources mixed by a rotation matrix. We theoretically analyze the proposed criterion and assess its performance through numerical simulations involving three distinct types of bounded signals. Our theoretical and experimental findings underscore the efficacy of the $\ell_\infty$ norm as a suitable contrast function for antisparse bounded sources, showcasing its superior performance relative to a state-of-the-art algorithm.
\end{abstract}


\section{Introduction}
\label{sec:intro}

Blind Source Separation (BSS) \cite{Comon2010} consists in recovering a set of signals, called sources, which have been combined by a mixing system. Its application covers a wide range of problems in different areas such as audio, speech, image processing, seismic reflection, communications, astronomy, chemistry, biomedicine, among others \cite{Comon2010}. In the case of linear, instantaneous and determined mixtures, we adopt the following model
\begin{equation}
    \mathbf{X} = \mathbf{H}\mathbf{S},
    \label{eq:mixture_equation}
\end{equation}
where $\mathbf{S} = [\mathbf{s}_1 \; \mathbf{s}_2 \; \cdots \; \mathbf{s}_N]^T \in \mathbb{R}^{N \times T}$ is the matrix containing the source vectors $$\mathbf{s}_i = [s_i(0) \cdots s_i(T-1)], \; \forall \; i=1, \; 2, \; \cdots, \; N,$$ with $T$ available samples; $\mathbf{X} \in \mathbb{R}^{N \times T}$ are the observed mixtures, and $\mathbf{H} \in \mathbb{R}^{N \times N}$ is the mixing system.

We can separate the sources by using a system $\mathbf{W} \in \mathbb{R}^{N \times N}$
\begin{equation}
    \mathbf{Y} = \mathbf{WX},
    \label{eq:separation_model}
\end{equation}
such that
\begin{equation}
    \mathbf{WH} = \mathbf{DP}.
\end{equation}
where $\mathbf{D}$ and $\mathbf{P}$ are diagonal and permutation matrices, respectively. In simple terms, a perfect source separation should recover the sources up to a scale ambiguity and an index permutation. Which means that the BSS problem admits an infinite number of possible solutions.


Typically, in Blind Source Separation (BSS), the objective is to recover prior information that has been attenuated during the mixing process. For many years, Independent Component Analysis (ICA) has been the go-to method. In the ICA paradigm, we assume that the sources are statistically independent, and we endeavor to optimize an appropriate contrast function to restore this property. Among the criteria for ICA, one can find mutual information, second- and higher-order statistics \cite{Fantinato2019}, nonlinear decorrelation \cite{Hyvarinen2001}, and information-theoretic approaches \cite{Moraes2021}. Nonetheless, in certain applications, the assumption of signal independence does not hold \cite{Friman2002, Bobin2015, Wu2017, Xie2020}, prompting the exploration of alternative priors.

Among the alternative priors for BSS, we have the Sparse Components Analysis (SCA) \cite{Xie2020, Gribonval2006, Georgiev2006, Naini2008, Elad2010, Duarte2011}, which explores the sparsity of the sources in some useful domain; the Disjoint Component Analysis (DCA) \cite{Anemuller2007, Nose-Filho2017}, which assumes that only one source is active at a given time instant; the Non-Negative Matrix Factorization (NMF) \cite{Cichocki2006, Cichocki2009, Hsieh2010}, which supposes that neither the source samples nor the mixing system coefficients take negative values; and, finally, the Bounded Component Analysis (BCA) \cite{Cruces2010, Erdogan2013, Inan2014}, which uses the sources bounded support as prior information. The latter prior is the focus of this paper.

Since the work \cite{Cruces2010}, BCA has been largely used across BSS problems: it can be used to separate non-negative sources \cite{Ouedraogo2014, Boulais2015}, and signed ones \cite{Erdogan2013, Inan2014, Cruces2015a, Cruces2015b} as well as dependent and independent sources \cite{Erdogan2013}. Also, such framework can be used for convolutive mixtures \cite{Inan2015}, and non-stationary sources \cite{Inan2015b}. 

In this work, we define a set of $N$ antisparse bounded sources, $\mathbf{S} = [\mathbf{s}_1 \; \mathbf{s}_2 \; \cdots \; \mathbf{s}_N]^T \in \mathbb{R}^{N \times T}$, as a set of $N$ signals with finite real valued amplitudes, $s_i(n) \in [-A, A], \; i=1, \; \cdots, \;N$, $n=0, \; \cdots, \;T-1$. In this context, we provide the following contributions to BCA:
\begin{enumerate}
    \item To propose a new criterion for BCA, in which we minimize the sum of the $\ell_\infty$ norm of the estimates, followed by an algorithm able to separate independent bounded sources mixed by any invertible matrix or separate correlated bounded sources mixed by a rotation matrix
    \item To present some interesting relationships between the proposed criterion and others used in BSS.
   \item To compare the performance of the proposed algorithm with a state-of-the-art algorithm.
\end{enumerate}

The paper is organized as follows: in Section \ref{sec:theoretical_background}, we review the BCA literature, presenting its theoretical underpinnings. In Section \ref{sec:proposed_BCA_criterion}, we derive a criterion, based on the $\ell_\infty$ norm, to separate bounded sources and present a separation algorithm based on Givens Rotation, suitable to optimize the proposed criterion. In Section \ref{sec:relationship_other_criteria}, we establish the  relationships between the proposed criterion and other ones commonly used in BSS. In Section \ref{sec:simulation_results}, we evaluate our approach by considering different experimental scenarios for bounded sources. Finally, in Section \ref{sec:conclusion} we state the main conclusions of the paper.

\section{Theoretical Background}{\label{sec:theoretical_background}}

The first works on BCA used the bounded support assumption alongside the sources independence one \cite{Mansour2002, Erdogan2006, Vrins2007}. The assumption of independent sources was set aside after the work of Cruces \cite{Cruces2010}, where the author stated three main conditions to recover bounded sources:

\begin{enumerate}

\item \textit{Compactness and Nondegeneracy of the Sources}: all sources are nondegenerate random variables of compact support.

\item \textit{Cartesian Decomposition of the Sources Convex Support}: the minimum convex hull of the joint support of the sources can be decomposed into the Cartesian product of the convex hull of each source support,

\begin{equation}
    \mathcal{S}_{\mathbf{S}} =  \mathcal{S}_{\mathbf{S}_1} \times \cdots \times \mathcal{S}_{\mathbf{S}_N}.
    \label{eq:joint_support_decomposition}
\end{equation}

\item \textit{Lossless Mixing}: the mixing matrix has full column-rank.
\end{enumerate}

Condition 2) represents an assumption about the sources that is less strict than the one based on statistical independence. If the sources are independent, their joint probability density function (pdf) can be decomposed into the product of the marginal ones; this also leads to Equation (\ref{eq:joint_support_decomposition}). On the other hand, if bounded sources satisfy (\ref{eq:joint_support_decomposition}), they are not necessarily independent \cite{Cruces2010}, \cite{Erdogan2013}. Such condition also implies that the extreme points of the sources joint support can be decomposed into the Cartesian product of the extreme points of the sources individual supports. This property is referred to as \textit{Extreme Points Condition} (EPC).

\begin{figure}[htb]
    \centering
    \includegraphics[scale=0.3]{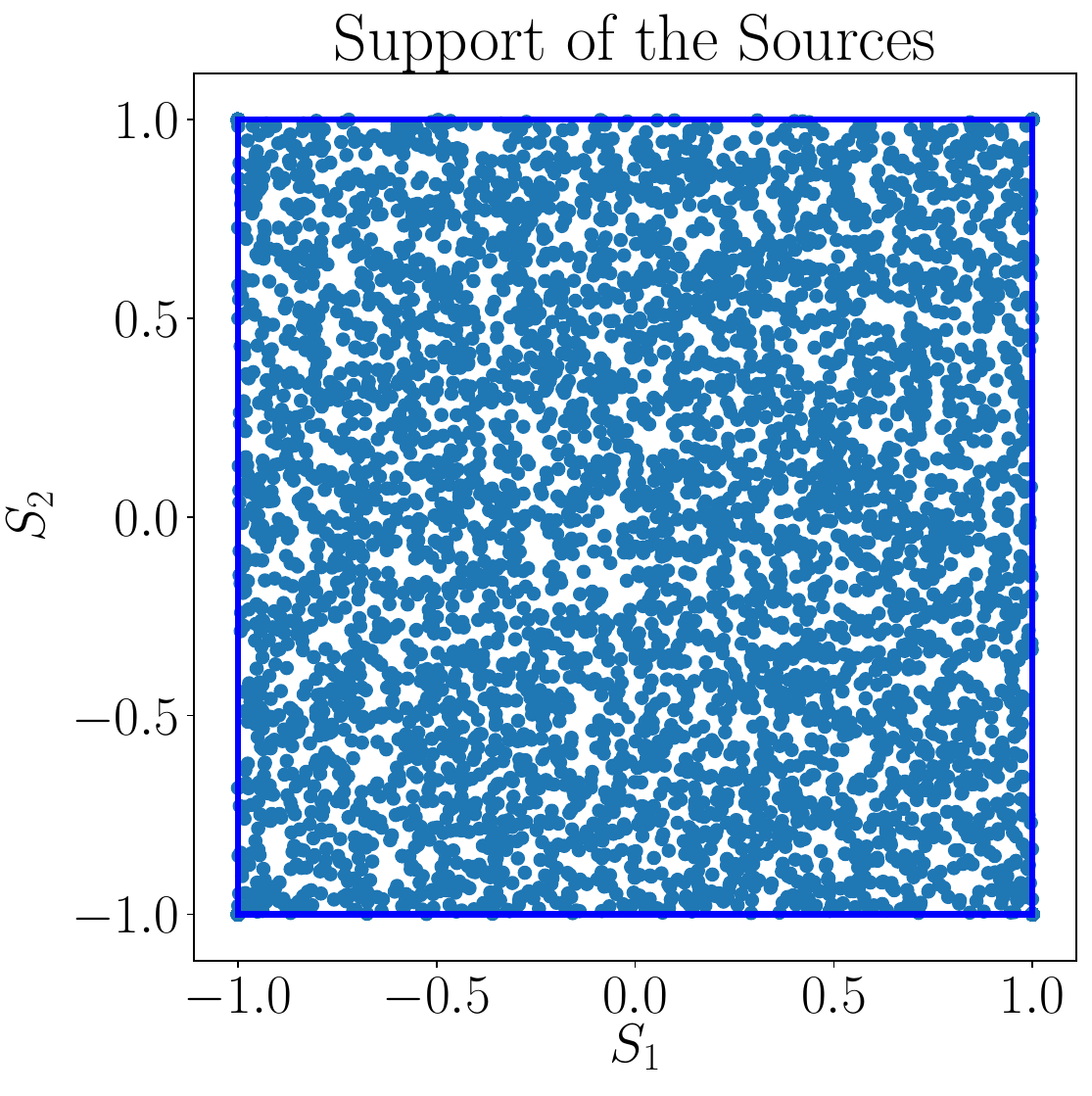}
    \caption{Illustration of the Extreme Points Condition of Cruces' work \cite{Cruces2010} for the case of having two real valued bounded sources.}
    \label{fig:square_support}
\end{figure}

To better illustrate it, let us consider the case of two real valued sources. For signals with symmetric support $[-1, 1]$, the Extreme Points Condition states that the joint support of the sources must have samples at the vertices of the square, \textit{i.e.}, it must contain the points $[1,1], [1,-1], [-1,1], [-1,-1]$, as illustrated by Fig. \ref{fig:square_support}. In other words, the square depicted in Fig. \ref{fig:square_support} is the smallest convex set that contains all of the source samples. If one can fit the estimate samples inside the smallest convex set, then the sources are separated. This is the idea behind the Minimum Perimeter Criterion, presented in \cite{Cruces2010} and detailed below.

\textit{Minimum Perimeter Criterion} (MPC): this criterion looks for a separation matrix that produces estimates with minimum perimeter convex hulls. To extract a source, we have
\begin{equation}
    \textrm{J}_{\textrm{MPC}}(\mathbf{b}) = \dfrac{\textrm{L}(\mathbf{b}^H \mathbf{X})}{||\mathbf{b||_2}},
\end{equation}
where $\mathbf{b}$ is the extraction vector, $||\cdot||_2$ is the $\ell_2$ norm and $\textrm{L}(\cdot)$ is the perimeter of the convex hull. Such perimeter is defined as the derivative of the Lebesgue measure of the estimates convex hull
\begin{equation}
    \textrm{L}(\mathbf{Y}) = \lim_{\epsilon \to 0^+} \dfrac{\textrm{V}_2(\mathcal{S}_{\mathbf{Y}} + \epsilon\mathcal{S}_B) - \textrm{V}_2(\mathcal{S}_{\mathbf{Y}})}{\epsilon},
\end{equation}
where $\mathcal{S}_B$ is the unit radius ball and $\textrm{V}_2(\cdot)$ is the Lebesgue measure in $\mathbb{R}^2$.

\begin{figure}[htb]
    \centering
    \includegraphics[scale=0.3]{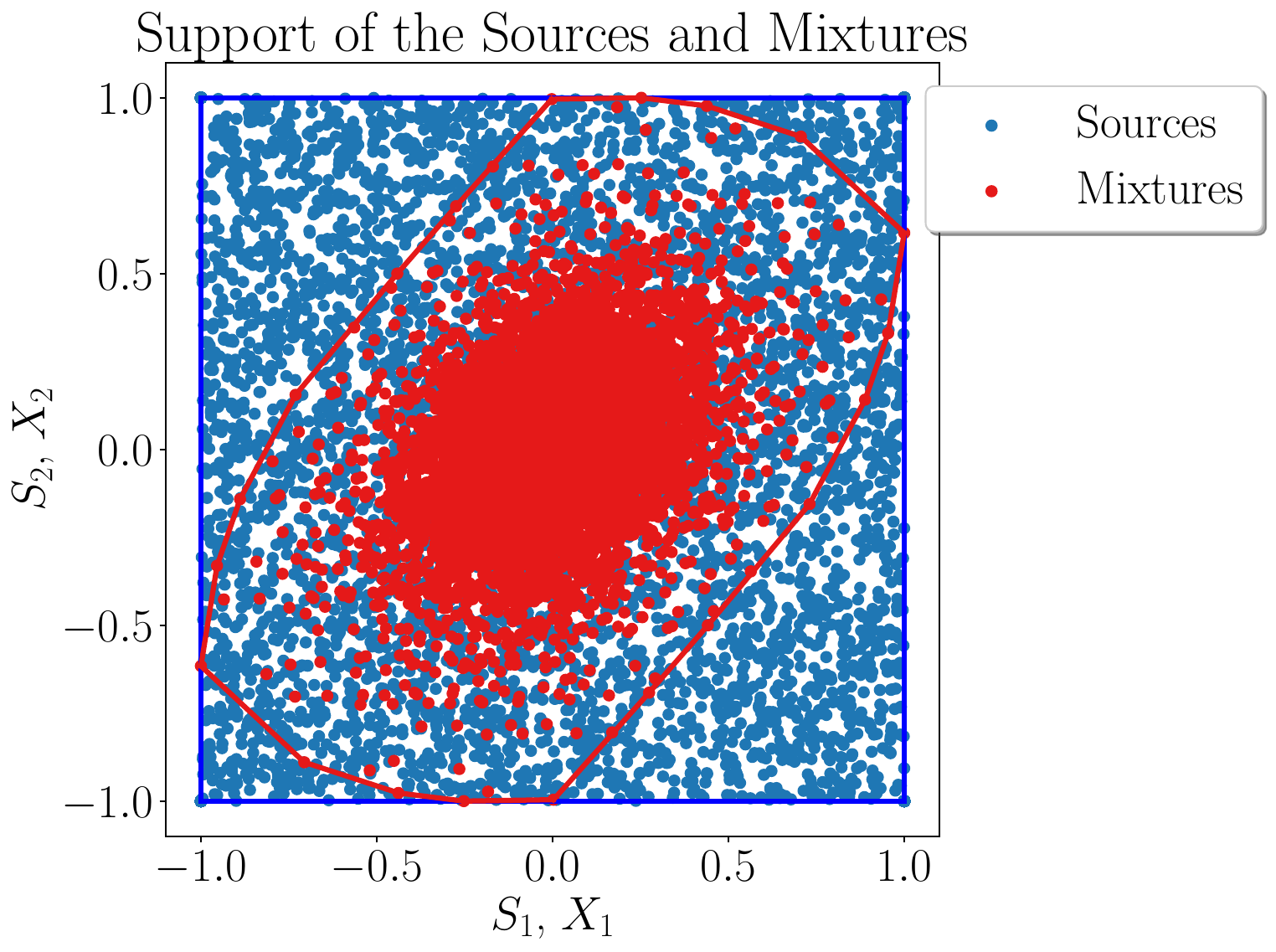}
    \caption{Sources and Mixtures Joint Support. Originally, the source samples lay inside the blue square. After the mixing, the mixture samples will be comprised inside the red ellipsoid, which is inside the original support}
    \label{fig:support_square_ellipsoid}
\end{figure}

Regarding the idea of a separation process based on support fitting, the work \cite{Erdogan2013} stated that after a linear mixing process, the mixtures samples will no longer fit a polytope, but instead, those samples will be comprised inside an ellipsoid. In Fig. \ref{fig:support_square_ellipsoid}, we illustrate this change in the support geometry, where the polytope is a square.

Fig. \ref{fig:support_square_ellipsoid} illustrates a very interesting property: compared to the square, the ellipsoid presents a smaller volume. Therefore, the maximization of the estimates' joint support volume, under the constraint that it must be inside a hypercube, is a strategy to recover the sources. This is precisely the key idea of the Volume Maximization criterion.

\textit{Volume Maximization} (VM): this criterion assumes that the sources have bounded supports, inscribed in an hypercube \cite{Erdogan2013}; after the mixing process, the mixtures' joint support is inscribed in an ellipsoid. The criterion, therefore, tries to maximize the estimates' joint support volume, subject to a hypercube fitting constraint. Algebraically, we have
\begin{equation}
    \textrm{J}_{\textrm{VM}}(\mathbf{W}) = K_N \dfrac{\sqrt{\textrm{det}(\mathbf{C}(\mathbf{Y}))}}{\Pi_{i=1}^{N} \mathcal{R}_i(\mathbf{Y})}
\end{equation}
where $\mathbf{C}(\mathbf{Y})$ is the cross-correlation matrix of the estimates, $\mathcal{R}_{i}(\mathbf{Y})$ is the range (the difference between the maximum and minimum value) of the $i$-th estimate, and $K_N = \frac{\pi^{N/2}}{\Gamma(\frac{N}{2} + 1)}$ is a constant.

\begin{figure}[htb]
\centering
    \includegraphics[scale=0.4]{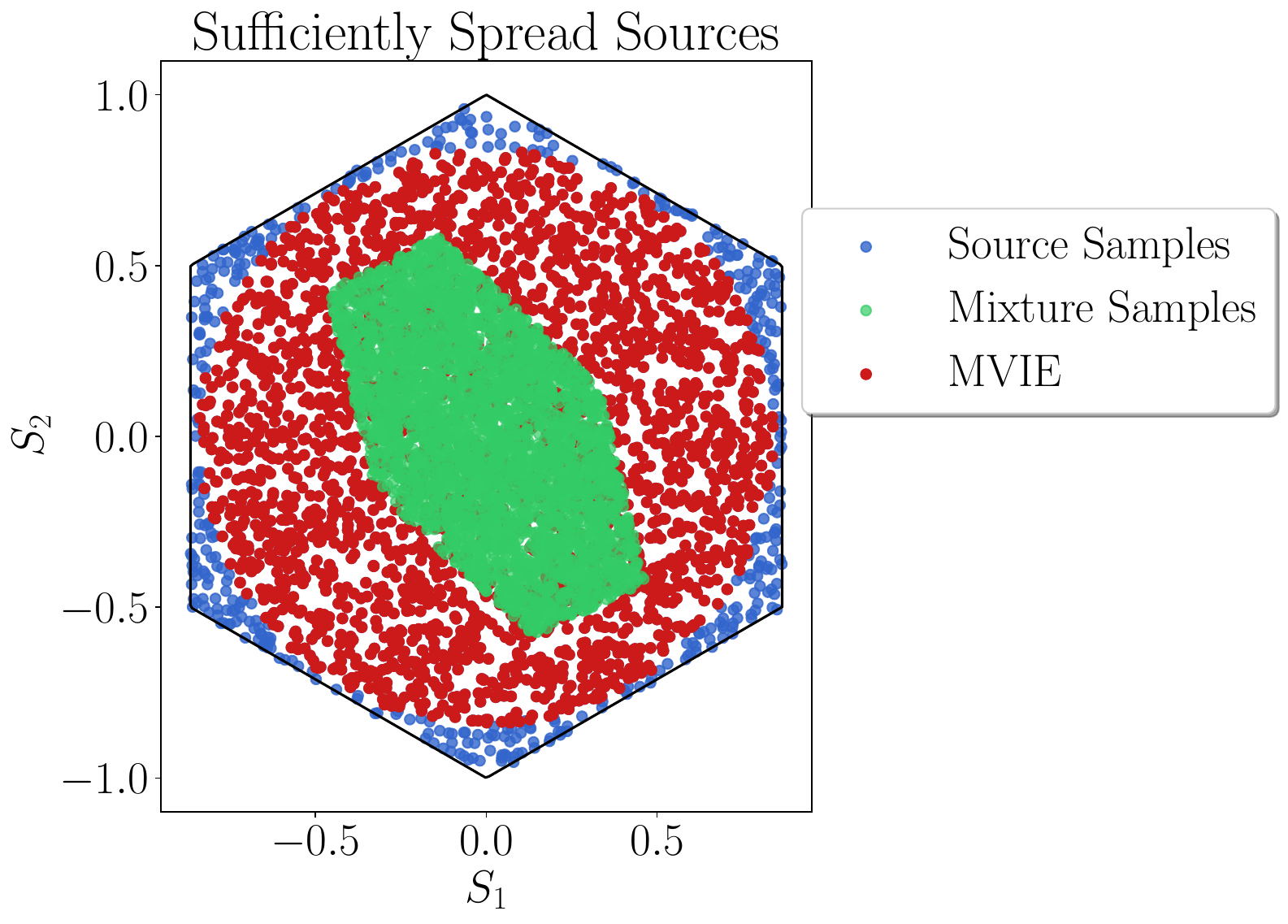}
    \caption{Sufficiently Spread Sources Condition. The sources support, represented by the blue hexagon, must contain the convex hull of the sources samples (represented by the blue dots), which in turn must contain the MVIE (red circle). After the mixing process, the mixture samples (green dots) also lay inside the MVIE.}
    \centering
    \label{fig:support_polytope_mvie}
\end{figure}

More recently, the works \cite{Tatli2021a, Tatli2021b} relaxed the Extreme Points Condition: it suffices that the source samples are sufficiently spread, such that the convex hull of their samples contains the Maximum Volume Inscribed Ellipsoid (MVIE). We illustrate this condition in Fig. \ref{fig:support_polytope_mvie}, considering two mixtures of two sources. Such works observed that the polytope must be sufficiently greater than the MVIE, and it also gives the prior information necessary to derive an adequate cost function to recover the sources. 

Having stated the theoretical background about BCA and presented its main results, next we propose an $\ell_\infty$ norm-based criterion to recover bounded sources.

\section{A $\ell_\infty$ norm-based Criterion and Method for BCA}\label{sec:proposed_BCA_criterion}

In \cite{Brotto2018}, we investigated the relationship between the $\ell_p$ norms, exploring different values of $p$, with some statistical features of signals. Regarding the $\ell_\infty$ norm, it corresponds to the Maximum Likelihood estimator for a signal uniformly distributed (and, therefore, with a bounded support). After that, we also successfully used such a norm in BSS, as is presented in the work \cite{Brotto2019}, in a more controlled and simpler scenario. 

Motivated by such results, in this section we present the proposed method based on finding a rotation matrix able to minimize the sum of the $\ell_\infty$ norm of the estimated sources, which can be summarized by the following optimization problem:
\begin{equation}
\begin{aligned}
& \underset{\mathbf{W}}{\text{minimize}}
& & J_{\infty}(\mathbf{W}) = \sum_{i=1}^{N}||\mathbf{y}_i||_\infty, \\
& \text{subject to}
& & \mathbf{WW^T=I}
\end{aligned}
\label{eq:optimization}
\end{equation}
For the case of having $N$ mixtures of $N$ independent bounded sources mixed by an invertible matrix, we can process the observed mixtures by a PCA-related technique (also known as a whitening process), leading to a set of signals that corresponds to the sources up to a rotation matrix \cite{Hyvarinen2001}, 
\begin{equation}
	\mathbf{\tilde{X}} = \mathbf{BX},
	\label{eq:whitening}
\end{equation}
where $\mathbf{B}$ is a whitening matrix. In this case $\mathbf{y}_i = \mathbf{w}_i^{T}\mathbf{\tilde{X}}$.

When we have $N$ mixtures of $N$ not necessarily independent bounded sources, mixed by an orthogonal matrix, we have that $\mathbf{y}_i = \mathbf{w}_i^{T}\mathbf{X}$. A proof demonstrating that the mixing process of bounded sources with any mixing matrix with unitary Frobenius norm will increase the $\ell_\infty$-norm of the sources, unless it is the identity matrix, is presented in \ref{sec:proof}. This result demonstrate that the minimization of the sum of the $\ell_\infty$-norm of the sources can be used for the estimation of an orthogonal separation matrix.

In the aforementioned cases, a rotation matrix $\mathbf{W}$ is a suitable structure for the separating system. It will produce estimates that correspond to the sources up to an index permutation, a scale ambiguity, and a sign inversion, which are natural in BSS.

Since $\mathbf{W}$ is an $N$-dimensional rotation matrix, it can be factorized into the product of $N(N-1)/2$ two-dimensional rotation matrices. Each one of those matrices performs a rotation in the $\mathbf{s}_i$-$\mathbf{s}_j$ plane, $i=1, 2, \cdots, N $, $j=i+1, \cdots, N$, \textit{i.e.}, the planes defined by each pair of sources. We can perform such procedure using Givens Rotations \cite{Golub1996}, \cite{Shah2018}, as detailed in Algorithm \ref{alg:givens_rotations}. 

\begin{algorithm}[htb!] 
\setstretch{1}
\caption{Givens Rotations}
\label{alg:givens_rotations}
\begin{algorithmic}
\Require 
$N$: Number of Sources; 
$\mathbf{X}$: Mixtures Data Matrix;
MAX\_ITER: Number of Maximum Iteration;
$\mu_0$: Initial Grid Search Step.
\Ensure $\mathbf{W}$: Optimal Rotation Matrix
    \State $\mu \gets \mu_0$
    \State $\mathbf{W} \gets \mathbf{I}_N$, $J_\textrm{min} \gets \infty$
    \For {$i \gets 1$ to  $MAX\_ITER$}
    \For{$m \gets 1$ to  $N-1$}
        \For{$n \gets m+1$ to  $N$}
        \State $\mathbf{T} \gets \mathbf{I}_N$
        \For{$\theta \gets 0$ to $\pi$, with  grid step $\mu$}
            \State $\mathbf{T}[m,m] = \mathbf{T}[n, n] = \cos(\theta)$;  $\mathbf{T}[m,n] = -\mathbf{T}[n, m] = -\sin(\theta)$
            \State $J \gets \textrm{calculate cost } \textrm{J}_{\infty}(\mathbf{TW})$
            \If{$J < J_\textrm{min}$}
            \State $J_\textrm{min} \gets J$, $\mathbf{T}^* \gets \mathbf{T}$
            \EndIf
        \EndFor
        \State $\mathbf{W} \gets \mathbf{T}^*\mathbf{W}$
        \EndFor    
    \EndFor
    \State $\mu \gets \mu/1.5$
    \EndFor \\
    \Return {$\mathbf{W}$}
\end{algorithmic}
\end{algorithm}

In Algorithm \ref{alg:givens_rotations}, we perform a grid search for the optimal rotation angle for all of the $\mathbf{s}_i$-$\mathbf{s}_j$ planes. Since we are searching the optimal value for an angle, we perform this search on the interval $[0,\pi)$. The Givens Rotation have been extensively used in BSS, specially when the cost function is not differentiable. Also, Algorithm \ref{alg:givens_rotations} has a computational complexity of $\mathcal{O}(MN^2g)$, where $M$ denotes the maximum number of iterations, $N$ the number of sources and $g$ the maximum number of points of the grid search. Since both $M$ and $g$ are constants, the proposed algorithm has computational complexity $\mathcal{O}(N^2)$.

Let us now present some relationships between our proposed $\ell_\infty$ norm criterion and others used in BSS.

\section{Relationships among the $\ell_{\infty}$ norm Criterion, the Lebesgue Measure and the Minimum Entropy Criterion}{\label{sec:relationship_other_criteria}}

In this section, we discuss the connections of the proposed criterion with the Lebesgue Measure and the Minimum Entropy Criterion. The Lebesgue measure can be seen as the extension to higher dimensions of the notions of length (1D), area (2D) and volume (3D). When applied to an interval $I = [a,b]$, the Lebesgue measure $\mu(I)$ corresponds to its length, $\mu(I) = b-a$. If a set $\mathcal{S}$ can be decomposed as $\mathcal{S} = \mathcal{S}_1 \times \mathcal{S}_2$, we have $\mu(\mathcal{S}) = \mu(\mathcal{S}_1).\mu(\mathcal{S}_2)$. Applying this property to a set of bounded sources and using Condition 2) of Section \ref{sec:theoretical_background}, we have
\begin{equation}
    \mu(\mathcal{S}_{\mathbf{S}}) = \mu(\mathcal{S}_{\mathbf{S}_1}).\mu(\mathcal{S}_{\mathbf{S}_2}). \cdots . \mu(\mathcal{S}_{\mathbf{S}_N})
\end{equation}

Since all of the sources take values in the interval $[-A, A]$, 
\begin{equation}
    \mu(\mathcal{S}_{\mathbf{S}}) = \Pi_{i=1}^{N}(2A) = 2^N\Pi_{i=1}^{N}A = 2^N\Pi_{i=1}^{N}||\mathbf{s}_i||_\infty.
    \label{eq:Lebesgue_measure_Ss}
\end{equation}

Equation (\ref{eq:Lebesgue_measure_Ss}) states a relationship between the $\ell_\infty$ norm of each source and the Lebesgue measure of the joint support. Applying the logarithm, we have
\begin{equation}
    \log(\mu(\mathcal{S}_\mathbf{S})) = N\log(2) + \sum_{i=1}^{N}\log(||\mathbf{s}_i||_\infty).
\end{equation}
Since the logarithm is a monotonic function, the minimization of $\sum_{i=1}^{N}\log(||\mathbf{s}_i||_\infty)$ is equivalent to the minimization of $\sum_{i=1}^{N}||\mathbf{s}_i||_\infty$. Therefore, the $\textrm{J}_\infty$ criterion is also a Minimum Lebesgue Measure one.

Because we are considering real sources, their Lebesgue measure can be written as

\begin{equation}
\mu(\mathbf{s}_i) = \max_n(s_i(n)) -  \min_n(s_i(n)) = 2||\mathbf{s}_i||_\infty.
\end{equation}

As the sources are supposed to fit an hypercube, $\mu(\mathbf{s}_i)$ is a measure of the length its sides. Hence, when we minimize the proposed criterion, we are also minimizing the perimeter of the support of the sources.

Finally, the entropy of a source $H(\mathbf{s}_i)$ is upper bounded by its $\ell_\infty$ norm \cite{Cover2006}:

\begin{equation}
    H(\mathbf{s}_i) \leq \log(2||\mathbf{s}_i||_\infty).
\end{equation}

So, the minimization $\textrm{J}_\infty$ is also equivalent to the Minimum Entropy criterion. 

Having discussed the theoretical aspects of $\textrm{J}_\infty$, in the next section we will evaluate the proposed method (which will be referred to as Linf) with numerical simulations.



\section{Numerical Simulations}{\label{sec:simulation_results}}


In this section, we assess the proposed approach under three distinct scenarios: (1) Digital Communications, (2) Gray Scale Images, and (3) Correlated Sources. Our method is contrasted with the Maximum Volume Inscribed Ellipsoid (referred to here as VM) proposed in \cite{Erdogan2013}, which employs the $\textrm{J}_{\textrm{MVIE}}$ cost function outlined in Section \ref{sec:theoretical_background}. Even though this method was first proposed in 2013, it has also been employed in more recent works, such as \cite{Tatli2021, Erdogan2022, Tatli2023}, being solid enough to promote an adequate comparison of performance.

\subsection{Digital Communication Signals}

For the Digital Communication scenario, we consider a 4-PAM modulation scheme \cite{Proakis2007}, with symbols $[\pm 3, \; \pm 1]$, $N = 2,\; 5, \; 10$ sources, with $T = 1000$ samples each. We consider two kinds of symbol distributions: uniform (equiprobable), such that $P(+3) = P(-3) = P(1) = P(-1) = 0.25$, and multimodal (non-equiprobable), with $P(+3) = P(-3) = 0.375$, and $P(-1) = P(1) = 0.125$; we also include the extreme points in these simulations. In digital communications, this condition does not pose any restriction: it could be achieved by adding to each message a data sequence containing the extreme points, like an error correction code. The difference between including the extreme points and sending a reference code is that in the case of the reference code it is necessary to know the position (time instant) that this code has been sent, while in the case of the extreme points, it is not.

\begin{figure}[H]
\begin{subfigure}{.3\textwidth}
  \includegraphics[width=1.1\textwidth]{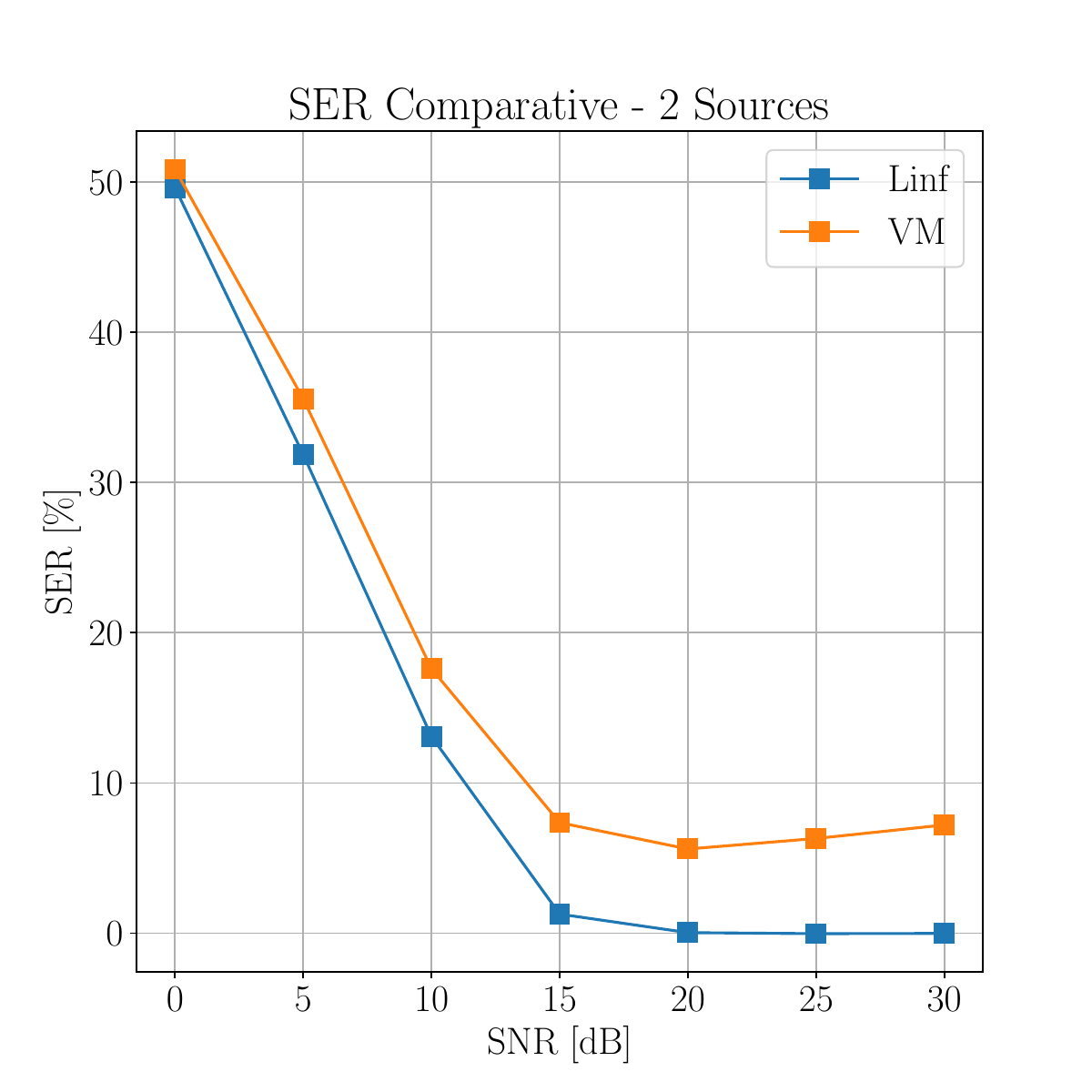}
  \caption{2 Sources}
  \label{fig:uniform-N-2}
  \centering
\end{subfigure}
\begin{subfigure}{.3\textwidth}
  \centering
  \includegraphics[width=1.1\textwidth]{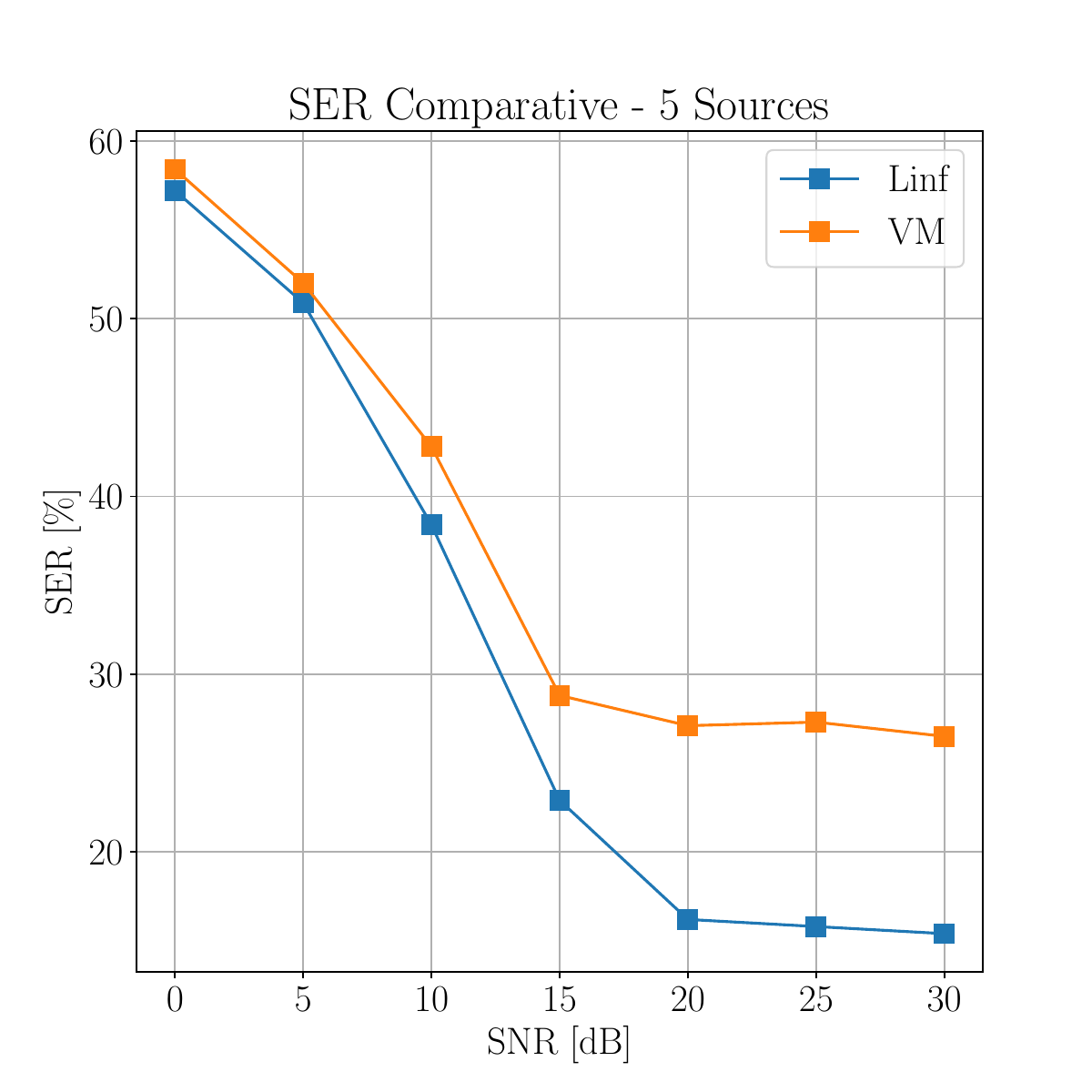}  
  \caption{5 Sources}
  \label{fig:uniform-N-5}
\end{subfigure}
\centering
\begin{subfigure}{.3\textwidth}
  \includegraphics[width=1.1\textwidth]{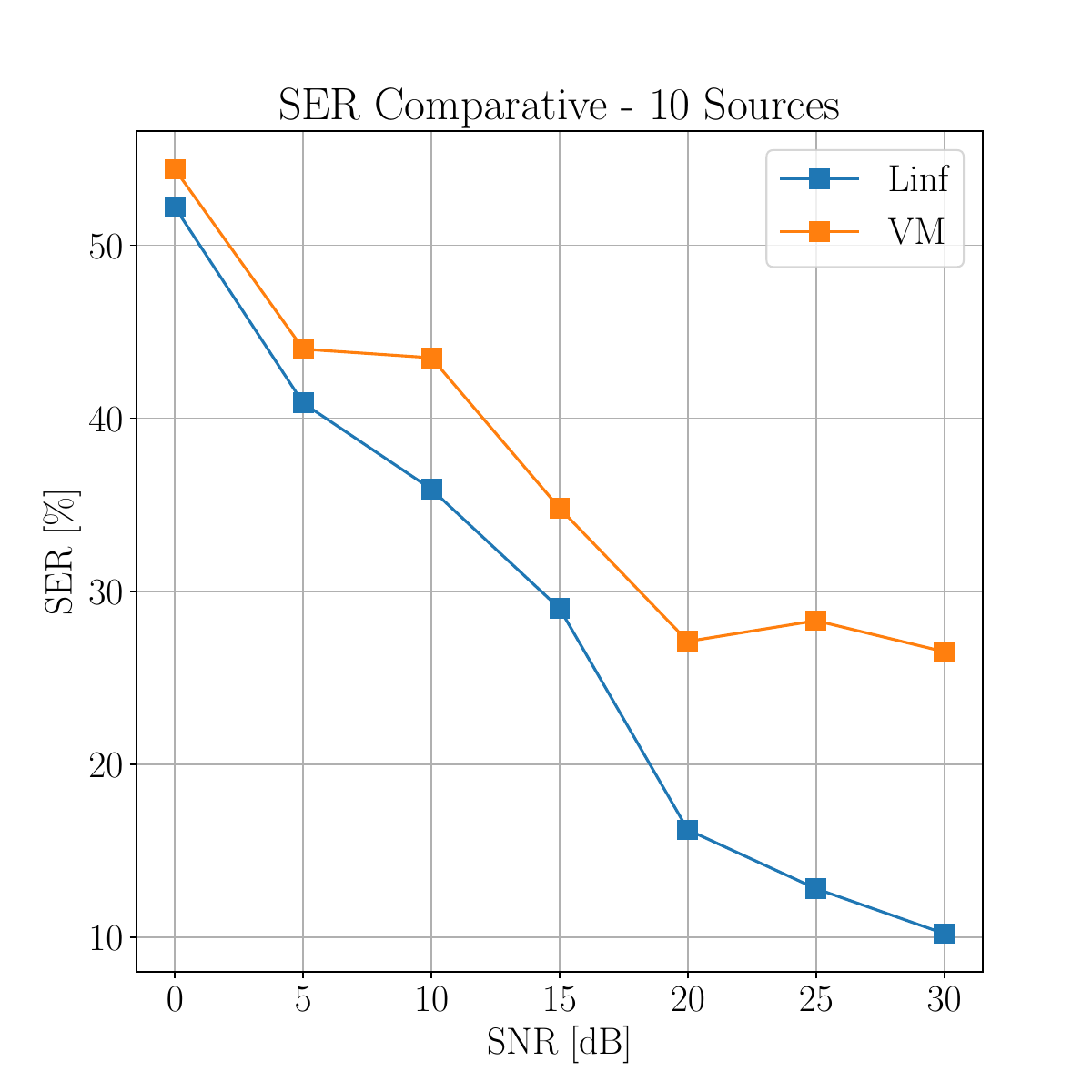}  
  \caption{10 Sources}
  \label{fig:uniform-N-10}
    \centering
\end{subfigure}
\centering
    \caption{SER Comparative - Uniform Distribution. Considering 2, 5 and 10 sources, we present the mean SER obtained by Linf (blue line) and VM (orange line) for SNRs = 0, 5, 10, 15, 20, 25 and 30 dB.}
    \label{fig:complete-uniform-ser}
\end{figure}

\begin{figure}[H]
\begin{subfigure}{.3\textwidth} 
  \includegraphics[width=1.1\textwidth]{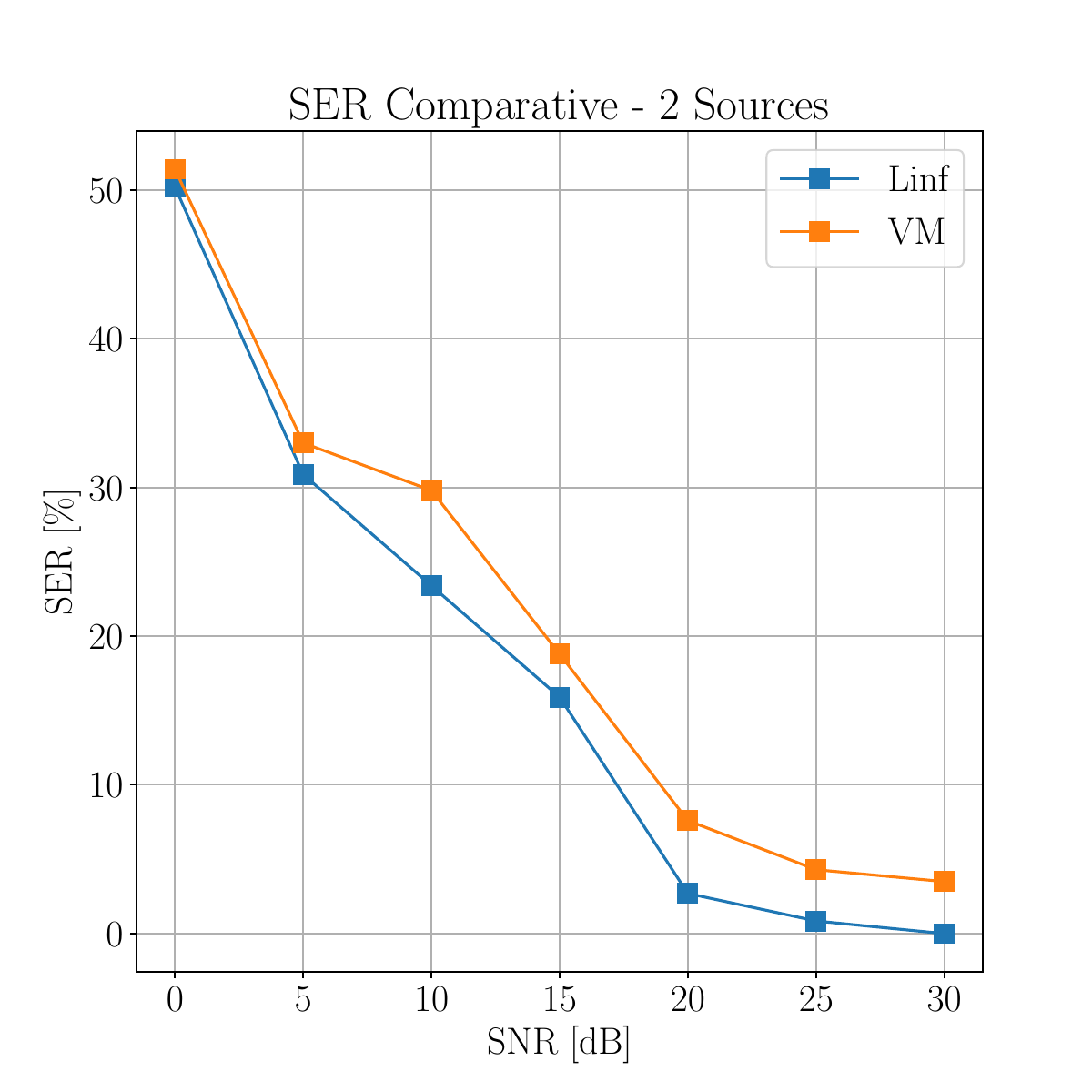}
  \caption{2 Sources}
  \label{fig:multimodal-N-2}
  \centering
\end{subfigure}
\begin{subfigure}{.3\textwidth}
  \centering
  \includegraphics[width=1.1\textwidth]{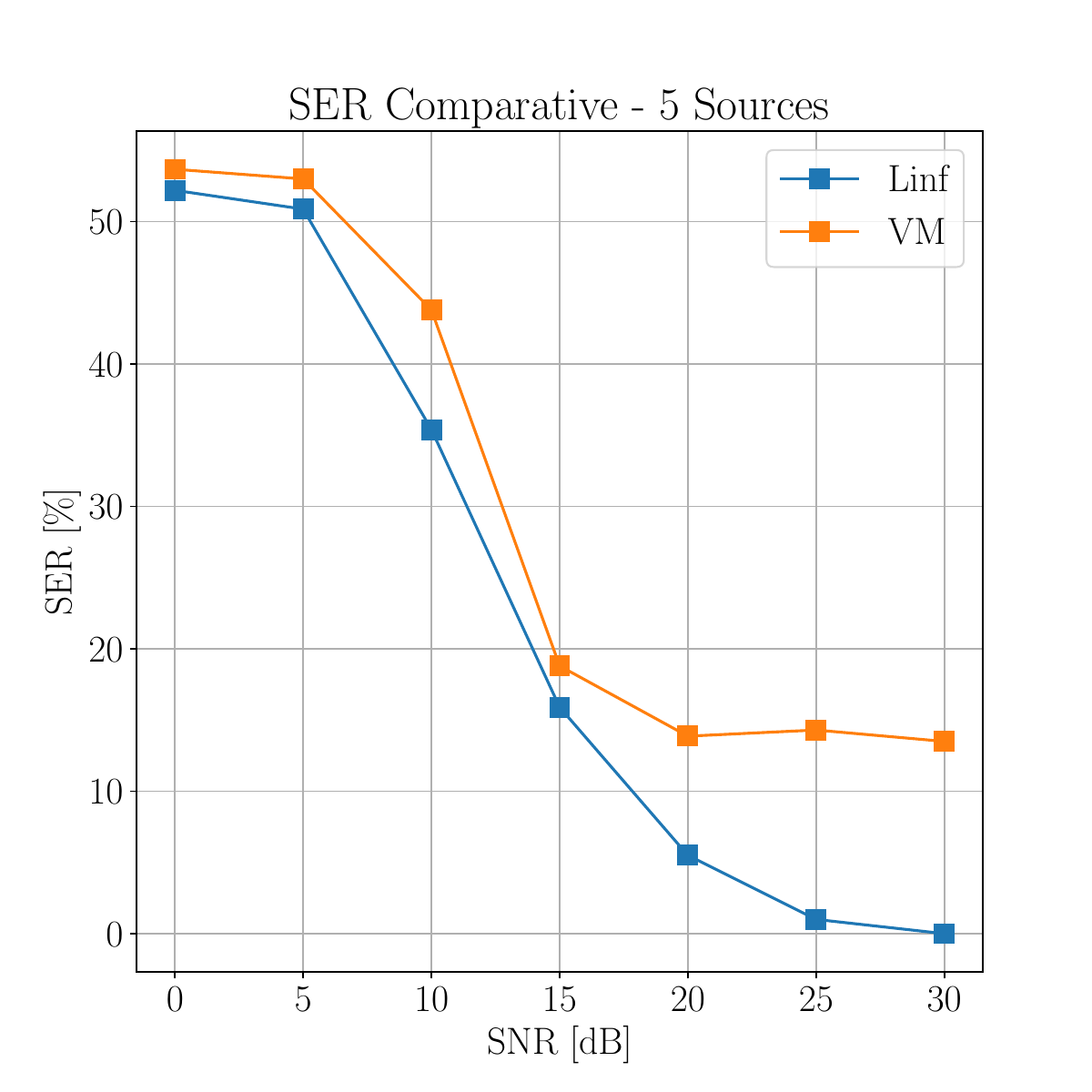}  
  \caption{5 Sources}
  \label{fig:multimodal-N-5}
\end{subfigure}
\centering
\begin{subfigure}{.3\textwidth}
  \includegraphics[width=1.1\textwidth]{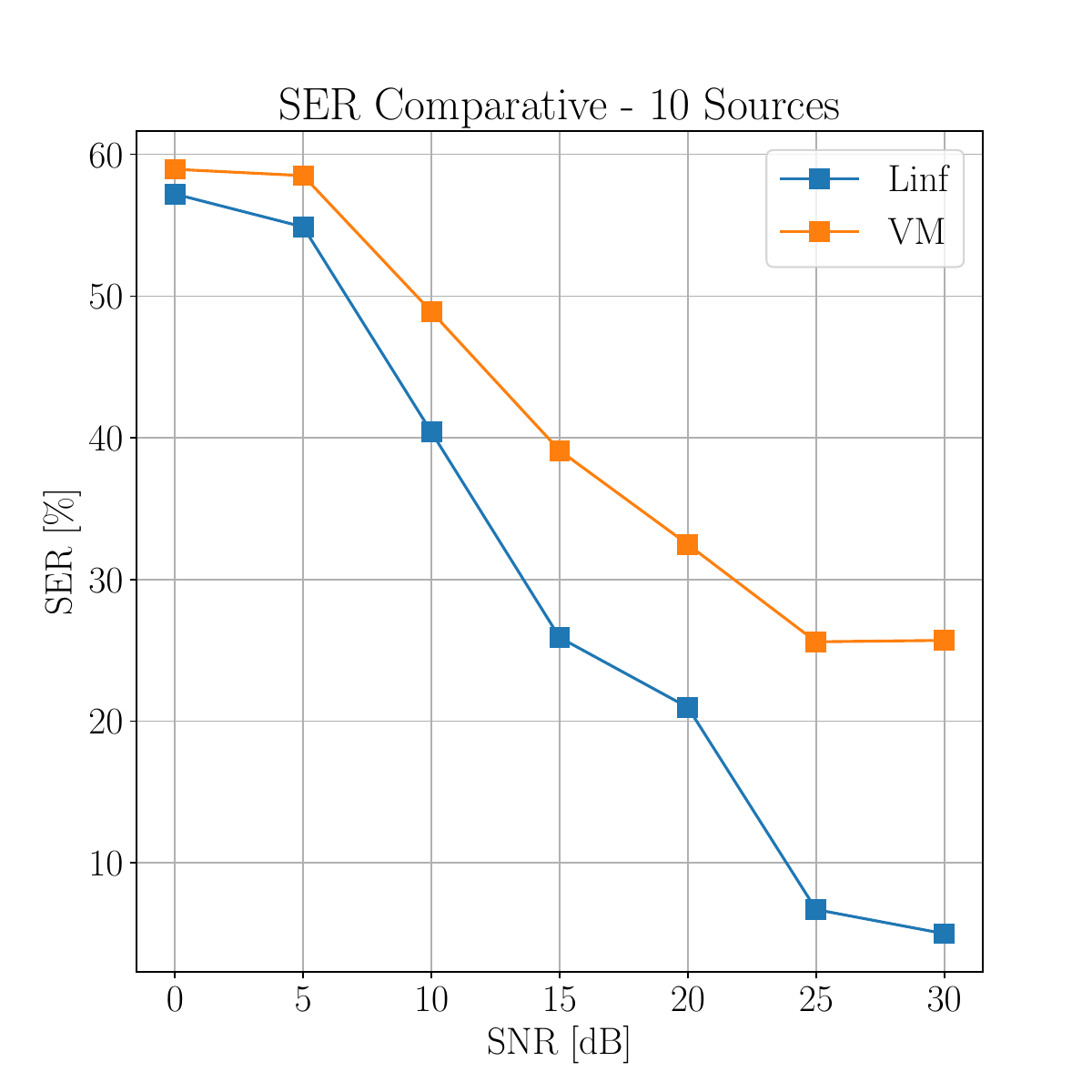}  
  \caption{10 Sources}
  \label{fig:multimodal-N-10}
    \centering
\end{subfigure}
\centering
    \caption{SER Comparative - Multimodal Distribution. Considering 2, 5 and 10 sources, we present the mean SER obtained by Linf (blue line) and VM (orange line) for SNRs = 0, 5, 10, 15, 20, 25 and 30 dB.}
    \label{fig:complete-multimodal-ser}
\end{figure}

For the mixing system, the coefficients are drawn from a $N(0,1)$ Gaussian distribution. To better simulate real world scenarios, we consider additive white Gaussian noise, for 7 levels of SNR (Signal-to-Noise Ratio): 0, 5, 10, 15, 20, 25 and 30 dB. After the separation, we apply the minimum Euclidean distance decision rule \cite{Proakis2007} to recover the symbols. To evaluate the performance of both methods, we adopt the Symbol Error Rate (SER) :
\begin{equation}
     \textrm{SER[\%]} = \dfrac{\textrm{\# incorrect symbols}}{\textrm{\# transmitted symbols}} \times 100.
\label{eq:SER}
\end{equation}
We present the obtained results in Figs. \ref{fig:complete-uniform-ser} (uniform distribution) and \ref{fig:complete-multimodal-ser} (multimodal one), considering 100 Monte Carlo simulations. 

We can observe that the performance of both methods decreases for low SNR and for a high number of sources, as was expected. It is interesting to note that the proposed method presents a mean performance that is superior to the one of the VM method for most of the SNRs considered, but both methods attained almost the same performance for a low SNR (5 and 0 dB). What stands out from Figs. \ref{fig:complete-uniform-ser} and \ref{fig:complete-multimodal-ser} is the fact that the distribution of the symbols played a minor role in terms of performance: the results are similar for both uniform and multimodal distributions. Broadly speaking, the information about the geometry of the support of the sources affected the performance more than the statistical information.

Next, we will present our simulation results for Gray Scale Images, a class of bounded signals in which we can not guarantee the Extreme Points Condition.

\subsection{Images}

\begin{figure*}
\centering
\begin{subfigure}{.24\textwidth}
  \centering
  \includegraphics[scale=0.22]{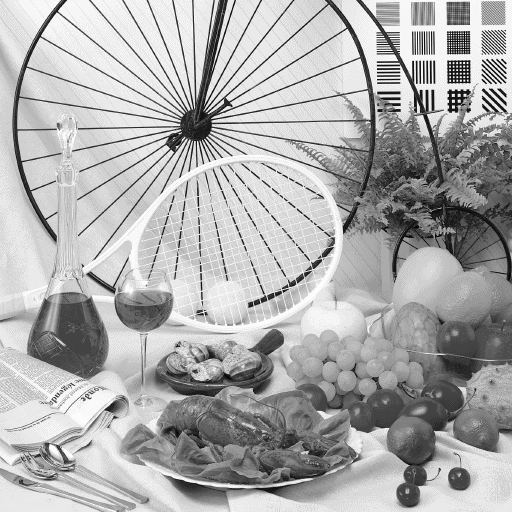}
  \caption{Source 1.}
  \label{fig:s1}
\end{subfigure}
\begin{subfigure}{.24\textwidth}
  \centering
  \includegraphics[scale=0.22]{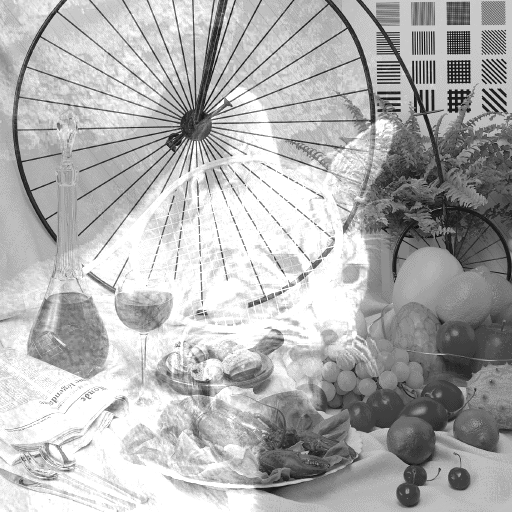}
  \caption{Mixture 1.}
  \label{fig:x1}
\end{subfigure}
\begin{subfigure}{.24\textwidth}
  \centering
  \includegraphics[scale=0.22]{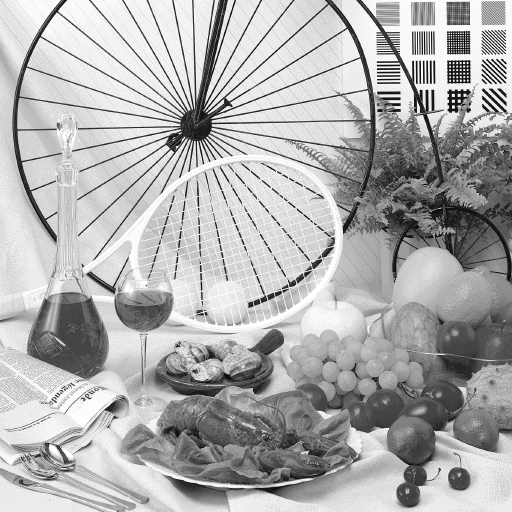}
  \caption{Estimate 1 (Linf).}
  \label{fig:y1_GR}
\end{subfigure}
\begin{subfigure}{.24\textwidth}
  \centering
  \includegraphics[scale=0.22]{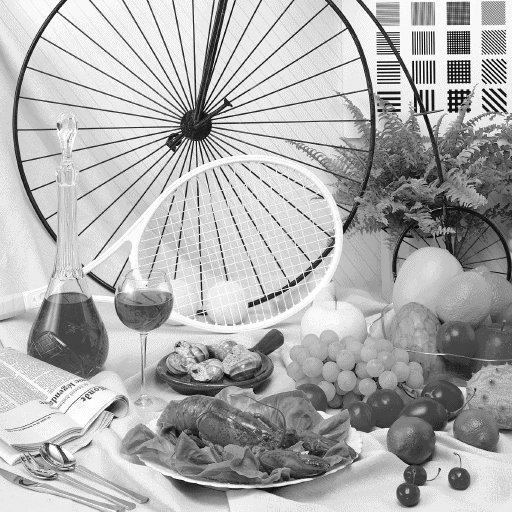}
  \caption{Estimate 1 (VM).}
  \label{fig:y1_Er}
\end{subfigure}

\begin{subfigure}{.24\textwidth}
  \centering
  \includegraphics[scale=0.22]{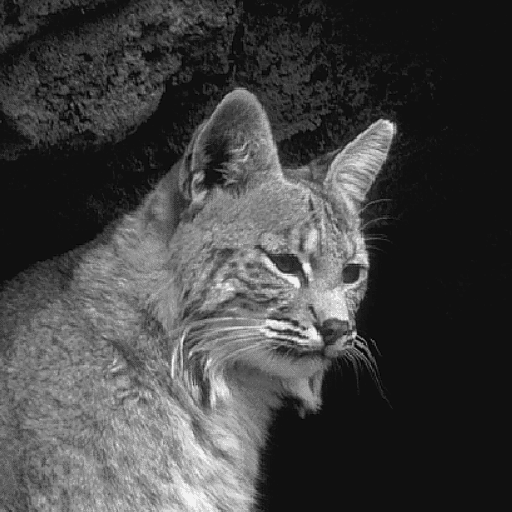}
  \caption{Source 2.}
  \label{fig:s2}
\end{subfigure}
\begin{subfigure}{.24\textwidth}
  \centering
  \includegraphics[scale=0.22]{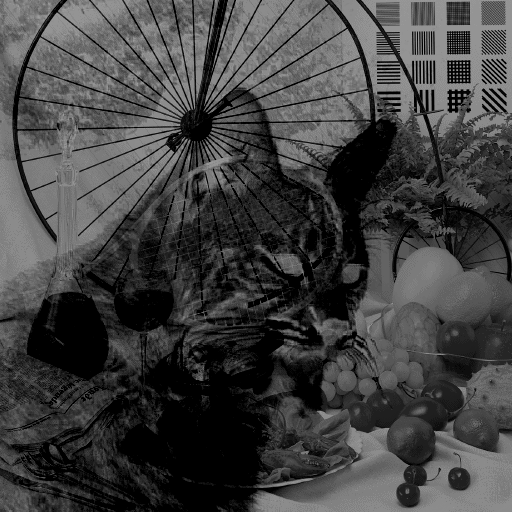}
  \caption{Mixture 2.}
  \label{fig:x2}
\end{subfigure}
\begin{subfigure}{.24\textwidth}
  \centering
  \includegraphics[scale=0.22]{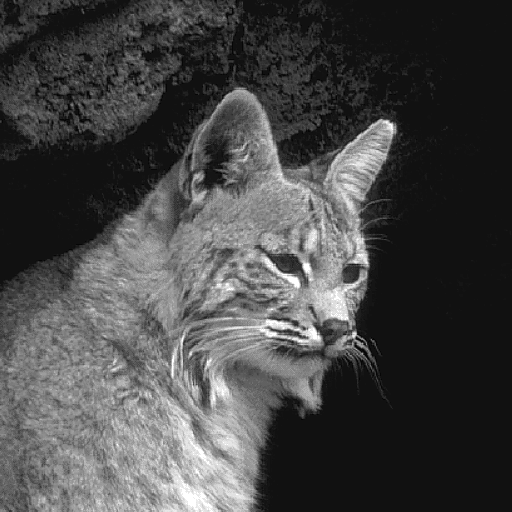}
  \caption{Estimate 2 (Linf).}
  \label{fig:y2_GR}
\end{subfigure}
\begin{subfigure}{.24\textwidth}
  \centering
  \includegraphics[scale=0.22]{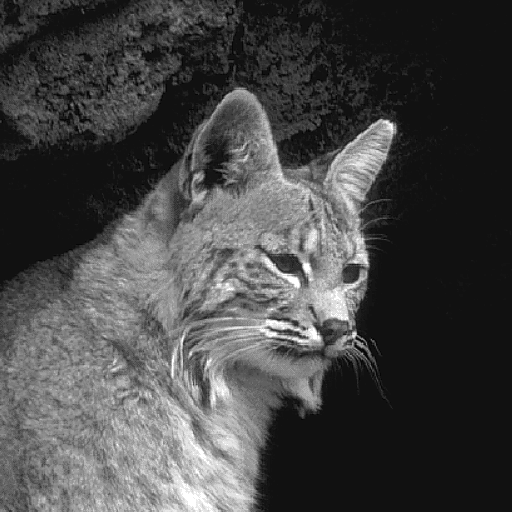}
  \caption{Estimate 2 (VM).}
  \label{fig:y2_Er}
\end{subfigure}
\caption{Results of a single simulation in a noiseless case. From left to right we present the sources, the mixtures, and the estimates obtained by the Linf and the VM algorithms.}
\label{fig:image_visualization}
\end{figure*}

The second scenario of simulation is in image processing, where we apply the proposed method to separate Gray Scale Images of dimensions 512 $\times$ 512 pixels. For the mixing system, the coefficients are drawn from a $N(0,1)$ Gaussian distribution and the sources correspond to the images illustrated in Figs. \ref{fig:s1} and \ref{fig:s2}, considering SNR = 0, 5, 10, 15, 20, 25 and 30 dB, for 10 Monte Carlo Simulations.

In Fig. \ref{fig:image_visualization} we present the obtained images throughout the BSS pipeline (sources-mixtures-estimates), for a single simulation in a noiseless scenario.

\begin{figure}[H]
    \centering
    \includegraphics[scale=0.5]{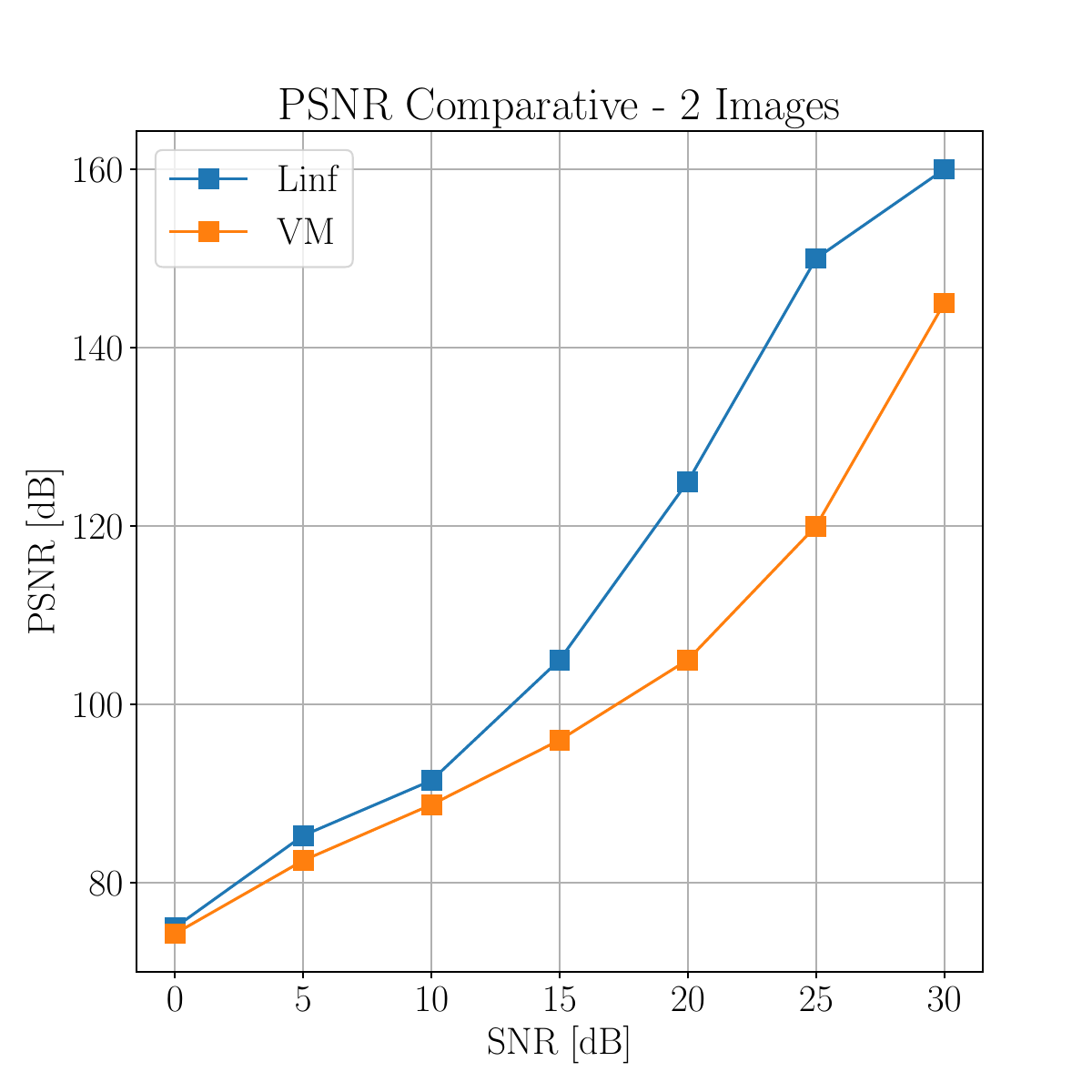}
    \caption{PSNR x SNR. We took the mean PNSR from 10 Monte Carlo simulations, considering SNR = 0, 5, 10, 15, 20, 25 and 30 dB, and we present the results obtained by the Linf method (blue line) and the VM one (orange line).}
    \label{fig:PSNR_SNR}
\end{figure}

We evaluate the methods with the Peak Signal-to-Noise Ratio (PNSR) \cite{Huynh-Thu2008}:
\begin{equation}
     \textrm{PSNR}(\mathbf{s}_i) = \underset{j}{\text{max}}\Bigg( 10.\textrm{log}\Bigg(  \dfrac{\underset{n}{\text{max }}s_i(n)^2}{\mathbb{E}\big[\big(s_i(n) - y_j(n)\big)^2\big]} \Bigg) \Bigg).
     \label{eq:PSNR}
 \end{equation}
The results for the 10 Monte Carlo simulation are presented in Figure \ref{fig:PSNR_SNR}. We can observe that both methods attained lower performance levels for lower SNRs, as was expected. We also note that the Linf method had a superior performance in most of the considered cases, but both methods attained a very similar performance for SNR = 5 and 0 dB. It is also important to note that even for SNR = 0 dB, both methods a high value of PNSR (arround 80 dB); one reason for this is that the PSNR is a best-scenario metric, since it compares the maximum value in the original sources to the error's mean energy. Therefore, even in a low SNR case, both methods attained an error's mean energy that was low enough compared to the maximuma pixel value of the originals images.

In addition to the performance obtained in this case and the vast practical usages of Gray Scale Images, the study of such a signal is interesting because we cannot guarantee the Extreme Points Condition. The good performance of the proposed method in this scenario shows that it is a suitable algorithm to recover bounded sources, even in the absence of such condition, corroborating the observations made in Section \ref{sec:theoretical_background}.

Finally, in the next section we will investigate how the proposed algorithm performs when dealing with correlated sources.

\subsection{Correlated Sources}


To obtain correlated sources, we start by generating independent uniform ones, taking values on the interval [-1, 1], and we apply the Copula-t technique, with 4 degrees of freedom \cite{Demarta2005}. As done in \cite{Erdogan2013} we adopt a Toeplitz structure to the sources cross-correlation matrix, whose first row is $[1 \; \rho \; \cdots \; \rho^{N-1}]$. Once again, we perform the separation for $N=2, \; 5$ and 10 sources, with $T = 1000$ samples, varying the correlation index $\rho$ from 0.0 to 0.9. Regarding the noise, we considered 3 SNR levels: 10, 20 and 30 dB. Since we are mainly interested in the effect of the correlation level, we added the extreme points $\{-1, 1\}^N$ to the sources. For the mixing systems, we considered orthogonal matrices randomly generated.

We evaluate both methods using the Intersymbol Interference (ISI) level of the global matrix $\mathbf{G}$:
\begin{equation}
    \textrm{ISI}(\mathbf{G}) \textrm{[dB]} = \dfrac{1}{N}\sum_{i=1}^{N}\textrm{ISI}(\mathbf{g}_i^{T}),
    \label{eq:ISI-G}
\end{equation}
\begin{equation}
     \textrm{ISI}(\mathbf{g}_i^{T})\textrm{[dB]} = 10 \textrm{log}_{10}  \dfrac{\sum_{l} |g_{i,l}|^2 - \max_l |g_{i,l}|^2}{\max_l |g_{i,l}|^2},
     \label{eq:ISI-gi}
\end{equation}
where $\mathbf{g}_i^{T}$ is the $i-$th row of $\mathbf{G}$. At a perfect source separation condition, $\mathbf{g}_i = \mathbf{e}_i$, and $\textrm{ISI}(\mathbf{G}) \to -\infty$; therefore, the lower  $\textrm{ISI}(\mathbf{G})$, the better the source separation. 

\begin{figure}[H]
\begin{subfigure}{.3\textwidth}
  \includegraphics[scale=0.27]{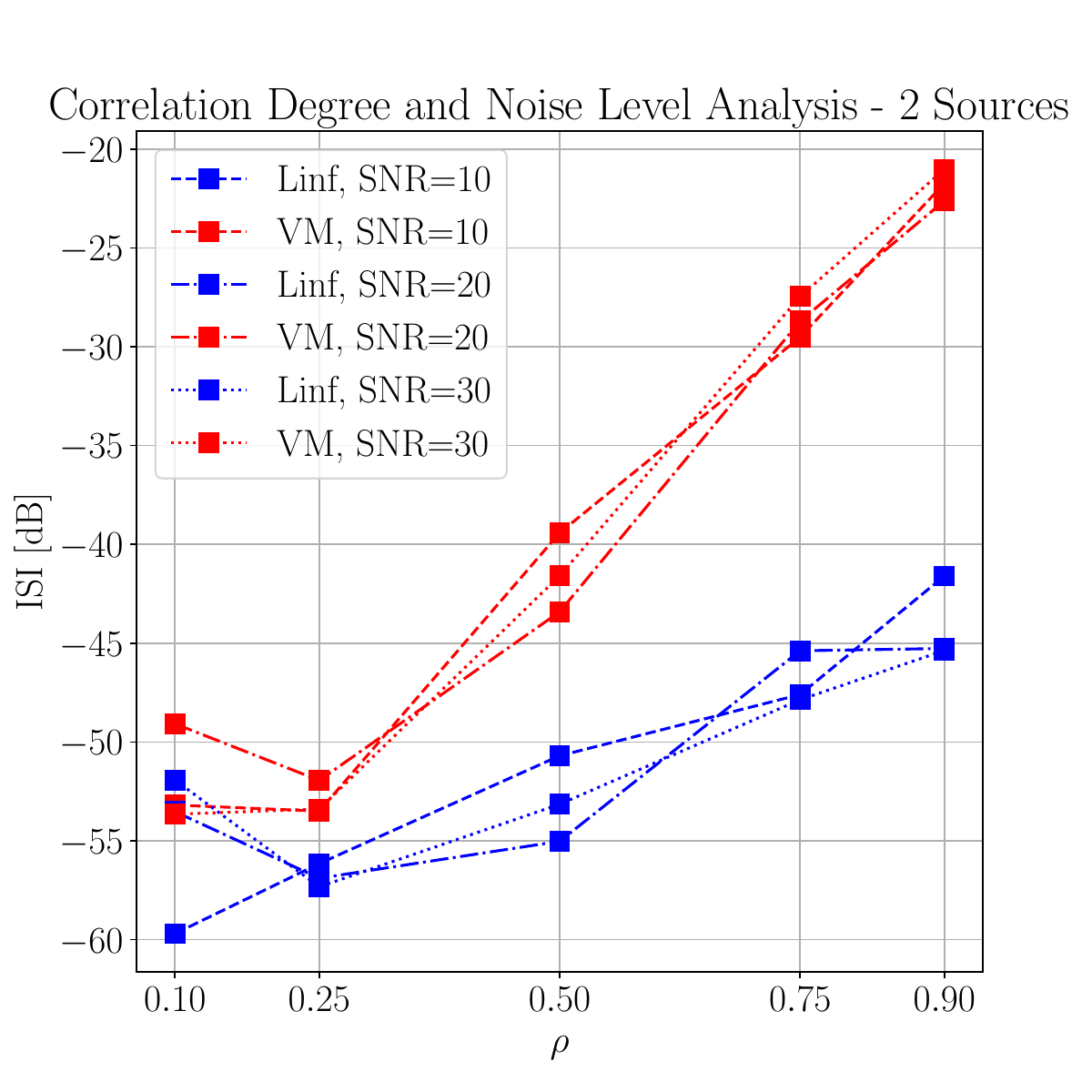}
  \caption{2 Sources}
  \label{fig:ISI-Inf-N-2}
  \centering
\end{subfigure}
\begin{subfigure}{.3\textwidth}
  \centering
  \includegraphics[scale=0.27]{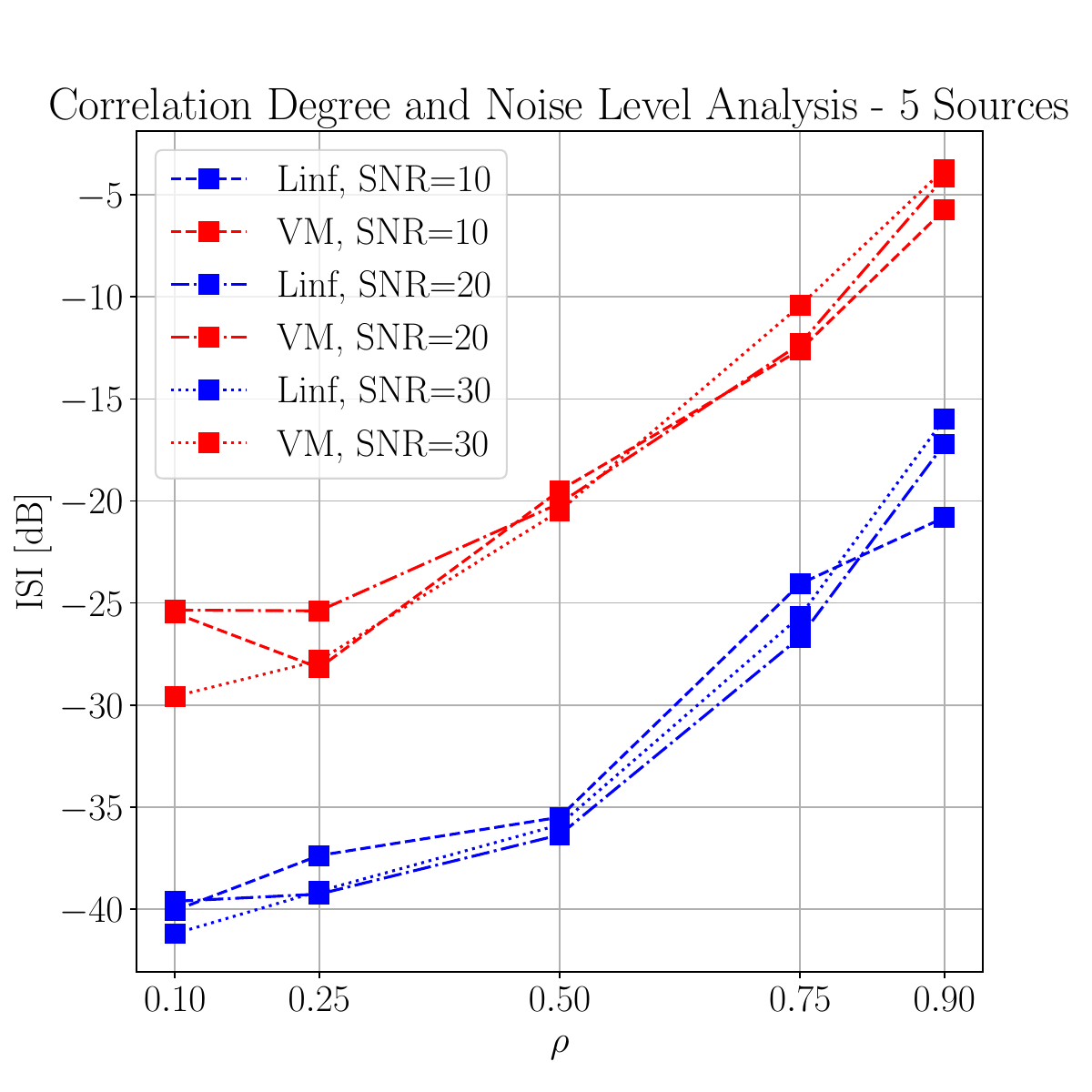}  
  \caption{5 Sources}
  \label{fig:ISI-Inf-N-5}
\end{subfigure}
\begin{subfigure}{.3\textwidth}
  \includegraphics[scale=0.27]{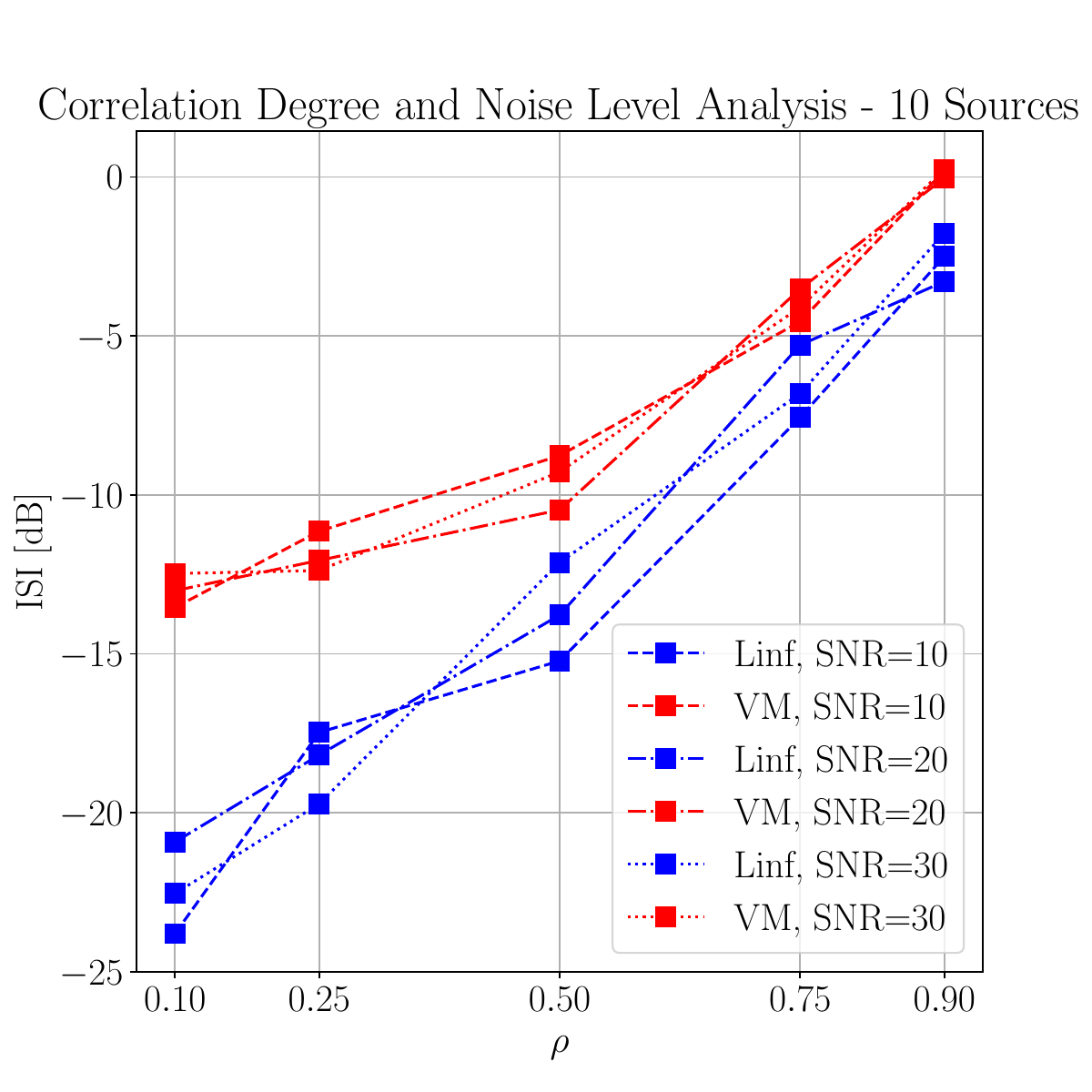}  
  \caption{10 Sources}
  \label{fig:ISI-Inf-N-10}
    \centering
\end{subfigure}
    \caption{Correlation and Noise Level Analysis. Considering 2, 5 and 10 sources, we compared the Linf method (blue line) with the VM one (red lines) for different noise levels SNR = 10 (dashed line), 20 (dash-dot line) and 30 (dotted line) dB and different correlation levels ($\rho=0.10, 0.25, 0.5, 0.75, 0.9$.}
    \label{fig:ISI-Inf}
\end{figure}

To investigate the effects of the correlation level, we present, in Figure \ref{fig:ISI-Inf}, the analysis of the correlation and noise levels, for 2, 5, and 10 sources, considering SNR= 10, 20 and 30 dB. Here, we considered 100 Monte Carlo simulations.

We can observe in Figure \ref{fig:ISI-Inf} that the more correlated the sources are the harder is to separate them, for both methods. It is also interesting to note that the correlation level affected more both methods' performance than the noise level. This is due to the fact that as we increase the correlation among the sources, we have less information to separate the sources, since they become more similar to each other. Another interesting fact, is that the Linf method had a superior performance in all the considered cases, making it an interesting option to deal with correlated sources, even when we have to separate a considerable number of signals (such as 10 sources, for exemple).

\section{Conclusion}{\label{sec:conclusion}}

In this work we presented a new BCA criterion for BSS. We stated sufficient conditions for the recovery of the sources and we showed that the $\ell_\infty$ norm is a contrast function for bounded signals, even when the condition of extreme points is not verified. 

We evaluated the quality of our approach with different kinds of signals, Digital Communications Messages, Gray Scale Images and Correlated Sources, each one posing different challenges. In all of the considered scenarios, the proposed method performed very well, \textit{i.e.} with a similar or better performance  than that obtained by the comparison method.

Since the proposed method led to very interesting results for correlated sources, in future works, we will investigate more general frameworks to cope with dependent signals.

\section{Acknowledgments}

The authors would like to thank the São Paulo Research Foundation (FAPESP) under grants \#2020/09838-0 (BI0S - Brazilian Institute of Data Science), \#2017/13025-2, \#2019/20899-4, and  CAPES, Code 001, for the financial support of this work.

\bibliographystyle{elsarticle-num}
\bibliography{myref}

@book{Johnson2012, place={Cambridge}, edition={2}, title={Matrix Analysis}, publisher={Cambridge University Press}, author={Horn, Roger A. and Johnson, Charles R.}, year={2012}}

@inproceedings{Tatli2023,
  title={A Bayesian Perspective for Determinant Minimization Based Robust Structured Matrix Factorization},
  author={Tatli, Gokcan and Erdogan, Alper T},
  booktitle={ICASSP 2023-2023 IEEE International Conference on Acoustics, Speech and Signal Processing (ICASSP)},
  pages={1--5},
  year={2023},
  organization={IEEE}
}

@inproceedings{Erdogan2022,
  title={An information maximization based blind source separation approach for dependent and independent sources},
  author={Erdogan, Alper T},
  booktitle={ICASSP 2022-2022 IEEE International Conference on Acoustics, Speech and Signal Processing (ICASSP)},
  pages={4378--4382},
  year={2022},
  organization={IEEE}
}

@article{Tatli2021,
  title={Polytopic matrix factorization: Determinant maximization based criterion and identifiability},
  author={Tatli, Gokcan and Erdogan, Alper T},
  journal={IEEE Transactions on Signal Processing},
  volume={69},
  pages={5431--5447},
  year={2021},
  publisher={IEEE}
}

@inproceedings{Brotto2019,
  title={Antisparse blind source separation},
  author={Brotto, Renan Del Buono and Nose-Filho, Kenji and Romano, Jo{\~a}o Marcos Travassos},
  booktitle={Proc. Signal Processing with Adaptive Sparse Structured Representations (SPARS) workshop},
  volume={52},
  year={2019}
}

@article{Brotto2018,
  title={Alternative criteria for predictive blind deconvolution},
  author={Brotto, Renan and Nose Filho, Kenji and Romano, Jo{\~a}o Marcos Travassos},
  journal={Journal of Communication and Information Systems},
  volume={33},
  number={1},
  year={2018}
}

@article{Moraes2021,
  title={Epanechnikov Kernel for PDF Estimation Applied to Equalization and Blind Source Separation},
  author={Moraes, Caroline PA and Fantinato, Denis G and Neves, Aline},
  journal={Signal Processing},
  volume={189},
  pages={108251},
  year={2021},
  publisher={Elsevier}
}

@article{Fantinato2019,
  title={A Second-Order Statistics Method for Blind Source Separation in Post-Nonlinear Mixtures},
  author={Fantinato, Denis G and Duarte, Leonardo T and Deville, Yannick and Attux, Romis and Jutten, Christian and Neves, Aline},
  journal={Signal Processing},
  volume={155},
  pages={63--72},
  year={2019},
  publisher={Elsevier}
}

@article{Wu2017,
title = {Blind Source Separation of Functional MRI Scans of the Human Brain based on Canonical Correlation Analysis},
journal = {Neurocomputing},
volume = {269},
pages = {220-225},
year = {2017},
issn = {0925-2312},
doi = {https://doi.org/10.1016/j.neucom.2017.01.106},
url = {https://www.sciencedirect.com/science/article/pii/S0925231217309967},
author = {Xingjie Wu and Ling-Li Zeng and Hui Shen and Ming Li and Yun-an Hu and Dewen Hu}
}

@article{Xie2020,
title = {Underdetermined Blind Separation of Source using lp-norm Diversity Measures},
journal = {Neurocomputing},
volume = {411},
pages = {259-267},
year = {2020},
issn = {0925-2312},
doi = {https://doi.org/10.1016/j.neucom.2020.06.029},
url = {https://www.sciencedirect.com/science/article/pii/S0925231220310006},
author = {Yuan Xie and Kan Xie and Shengli Xie}
}

@article{Friman2002,
title = {Exploratory fMRI Analysis by Autocorrelation Maximization},
journal = {NeuroImage},
volume = {16},
number = {2},
pages = {454-464},
year = {2002},
issn = {1053-8119},
doi = {https://doi.org/10.1006/nimg.2002.1067},
url = {https://www.sciencedirect.com/science/article/pii/S1053811902910670},
author = {Ola Friman and Magnus Borga and Peter Lundberg and Hans Knutsson}
}

@ARTICLE{Bobin2015,
  author={Jérôme Bobin and Jérémy Rapin and Anthony Larue and Jean-Luc Starck},
  journal={IEEE Transactions on Signal Processing}, 
  title={Sparsity and Adaptivity for the Blind Separation of Partially Correlated Sources}, 
  year={2015},
  volume={63},
  number={5},
  pages={1199-1213},
  doi={10.1109/TSP.2015.2391071}}

@ARTICLE{Vrins2007,
  author={Frdric Vrins and John A. Lee and Michel Verleysen},
  journal={IEEE Transactions on Neural Networks}, 
  title={A Minimum-Range Approach to Blind Extraction of Bounded Sources}, 
  year={2007},
  volume={18},
  number={3},
  pages={809-822},
  doi={10.1109/TNN.2006.889941}}

@ARTICLE{Ouedraogo2014,
  author={Wendyam Serge Boris Ouedraogo and Antoine Souloumiac and Mériem Jaïdane and Christian Jutten},
  journal={IEEE Transactions on Signal Processing}, 
  title={Non-Negative Blind Source Separation Algorithm Based on Minimum Aperture Simplicial Cone}, 
  year={2014},
  volume={62},
  number={2},
  pages={376-389},
  doi={10.1109/TSP.2013.2287683}}

@INPROCEEDINGS{Boulais2015,
  author={Axel Boulais and Yannick Deville and Olivier Bern},
  booktitle={2015 IEEE International Workshop of Electronics, Control, Measurement, Signals and their Application to Mechatronics (ECMSM)}, title={A geometrical blind separation method for unconstrained-sum locally dominant sources}, 
  year={2015},
  volume={},
  number={},
  pages={1-6},
  doi={10.1109/ECMSM.2015.7208711}}

@ARTICLE{Inan2015b,
  author={Huseyin A. Inan and Alper T. Erdogan},
  journal={IEEE Transactions on Signal Processing}, 
  title={A Convolutive Bounded Component Analysis Framework for Potentially Nonstationary Independent and/or Dependent Sources}, 
  year={2015},
  volume={63},
  number={1},
  pages={18-30},
  doi={10.1109/TSP.2014.2367472}}

@article{Mansour2002,
title = {Blind multiuser separation of instantaneous mixture algorithm based on geometrical concepts},
journal = {Signal Processing},
volume = {82},
number = {8},
pages = {1155-1175},
year = {2002},
issn = {0165-1684},
doi = {https://doi.org/10.1016/S0165-1684(02)00250-5},
url = {https://www.sciencedirect.com/science/article/pii/S0165168402002505},
author = {A. Mansour and N. Ohnishi and C.G. Puntonet},
}

@ARTICLE{Inan2015,
  author={Huseyin A. Inan and Alper T. Erdogan},
  journal={IEEE Transactions on Neural Networks and Learning Systems}, 
  title={Convolutive Bounded Component Analysis Algorithms for Independent and Dependent Source Separation}, 
  year={2015},
  volume={26},
  number={4},
  pages={697-708},
  doi={10.1109/TNNLS.2014.2320817}}

@ARTICLE{Erdogan2006,
  author={A.T. Erdogan},
  journal={IEEE Transactions on Signal Processing}, 
  title={A simple geometric blind source separation method for bounded magnitude sources}, 
  year={2006},
  volume={54},
  number={2},
  pages={438-449},
  doi={10.1109/TSP.2005.861800}}

@ARTICLE{Cruces2015b,
  author={Sergio Cruces},
  journal={IEEE Transactions on Signal Processing}, 
  title={Bounded Component Analysis of Noisy Underdetermined and Overdetermined Mixtures}, 
  year={2015},
  volume={63},
  number={9},
  pages={2279-2294},
  doi={10.1109/TSP.2015.2404312}}

@INPROCEEDINGS{Tatli2021b,
  author={Gokcan Tatli and Alper T. Erdogan},
  booktitle={ICASSP 2021 - 2021 IEEE International Conference on Acoustics, Speech and Signal Processing (ICASSP)}, 
  title={Generalized Polytopic Matrix Factorization}, 
  year={2021},
  volume={},
  number={},
  pages={3235-3239},
  doi={10.1109/ICASSP39728.2021.9413709}}

@ARTICLE{Tatli2021a,
  author={Gokcan Tatli and Alper T. Erdogan},
  journal={IEEE Transactions on Signal Processing}, 
  title={Polytopic Matrix Factorization: Determinant Maximization Based Criterion and Identifiability}, 
  year={2021},
  volume={69},
  number={},
  pages={5431-5447},
  doi={10.1109/TSP.2021.3112918}}

@ARTICLE{Cruces2015a,
  author={Sergio Cruces and Iván Durán-Díaz},
  journal={IEEE Transactions on Neural Networks and Learning Systems}, 
  title={The Minimum Risk Principle That Underlies the Criteria of Bounded Component Analysis}, 
  year={2015},
  volume={26},
  number={5},
  pages={964-981},
  doi={10.1109/TNNLS.2014.2329318}}

@INPROCEEDINGS{Hsieh2010,
  author={Hsin-Lung Hsieh and Jen-Tzung Chien},
  booktitle={2010 IEEE International Conference on Acoustics, Speech and Signal Processing}, 
  title={A new nonnegative matrix factorization for independent component analysis}, 
  year={2010},
  volume={},
  number={},
  pages={2026-2029},
  doi={10.1109/ICASSP.2010.5494945}}

@INPROCEEDINGS{Cichocki2006,
  author={A. Cichocki and R. Zdunek and S. Amari},
  booktitle={2006 IEEE International Conference on Acoustics Speech and Signal Processing Proceedings}, 
  title={New Algorithms for Non-Negative Matrix Factorization in Applications to Blind Source Separation}, 
  year={2006},
  volume={5},
  number={},
  pages={V-V},
  doi={10.1109/ICASSP.2006.1661352}}

@BOOK{Cichocki2009,
title = {Nonnegative Matrix and Tensor Factorizations: Applications to Exploratory Multi-way Data Analysis and Blind Source Separation},
publisher = {Wiley},
edition = {1},
year = {2009},
author = {Andrzej Cichocki and Rafal Zdunek and Anh Huy Phan and Shun-ichi Amari}
}

@BOOK{Elad2010,
title = {Sparse and Redundant Representations - From Theory to Applications in Signal and Image Processing},
publisher = {Springer, New York, NY},
edition = {1},
year = {2010},
author = {Michael Elad}
}

@ARTICLE{Georgiev2006,
  author={P. Georgiev and F. Theis and A. Cichocki},
  journal={IEEE Transactions on Neural Networks}, 
  title={Sparse component analysis and blind source separation of underdetermined mixtures}, 
  year={2005},
  volume={16},
  number={4},
  pages={992-996},
  doi={10.1109/TNN.2005.849840}}

@article{Naini2008,
title = {Estimating the mixing matrix in Sparse Component Analysis (SCA) based on partial k-dimensional subspace clustering},
journal = {Neurocomputing},
volume = {71},
number = {10},
pages = {2330-2343},
year = {2008},
note = {Neurocomputing for Vision Research Advances in Blind Signal Processing},
issn = {0925-2312},
doi = {https://doi.org/10.1016/j.neucom.2007.07.035},
author = {Farid {Movahedi Naini} and G. {Hosein Mohimani} and Massoud Babaie-Zadeh and Christian Jutten},
}

@ARTICLE{Demarta2005,
author={Stefano Demarta  and Alexander J. McNeil},
journal={International Statistical Review / Revue Internationale De Statistique}, 
title={The t Copula and Related Copulas.},
year={2005},
volume={73}, 
number={1},
pages={111-129},
}

@ARTICLE{Huynh-Thu2008,
author={Q. Huynh-Thu   and  M. Ghanbari},
journal={Electronics Letters}, 
title={Scope of validity of PSNR in image/video quality assessment},
year={2008},
volume={44}, 
number={13},
pages={800-801},
doi={10.1049/el:20080522}}

@ARTICLE{Erdogan2013,
  author={Alper T. Erdogan},
  journal={IEEE Transactions on Signal Processing}, 
  title={A Class of Bounded Component Analysis Algorithms for the Separation of Both Independent and Dependent Sources}, 
  year={2013},
  volume={61},
  number={22},
  pages={5730-5743},
  doi={10.1109/TSP.2013.2280115}}

@INPROCEEDINGS{Inan2014,
  author={Inan, Huseyin A. and Erdogan, Alper T.},
  booktitle={2014 48th Asilomar Conference on Signals, Systems and Computers}, 
  title={An extended family of bounded component analysis algorithms}, 
  year={2014},
  volume={},
  number={},
  pages={442-445},
  doi={10.1109/ACSSC.2014.7094481}}

@BOOK{Foucart2013,
title = {A Mathematical Introduction to Compressive Sensing},
publisher = {Springer},
edition = {1},
year = {2013},
author = {Simon Foucart and Holger Rauhut}
}

@ARTICLE{Cruces2010,
  author={S. {Cruces}},
  journal={IEEE Transactions on Signal Processing}, 
  title={Bounded Component Analysis of Linear Mixtures: A Criterion of Minimum Convex Perimeter}, 
  year={2010},
  volume={58},
  number={4},
  pages={2141-2154},}

@ARTICLE{Shah2018,
  author={S. A. W. {Shah} and K. {Abed-Meraim} and T. Y. {Al-Naffouri}},
  journal={IEEE Transactions on Signal Processing}, 
  title={Blind Source Separation Algorithms Using Hyperbolic and Givens Rotations for High-Order QAM Constellations}, 
  year={2018},
  volume={66},
  number={7},
  pages={1802-1816},}

@BOOK{Proakis2007,
title = {Digital Communications},
publisher = {McGraw-Hill Education, 5th ed.},
year = {2007},
author = {John Proakis and Masoud Salehi}
}

@BOOK{Cover2006,
	title = {Elements of Information Theory},
	publisher = {Wiley-Interscience},
	year = {2006},
	author = {Thomas. M. Cover, Joy A. Thomas}
}

@INCOLLECTION{Anemuller2007,
  author = {Anemuller, J.},
  title = {Maximization of component disjointness: a criterion for blind source
	separation},
  booktitle = {Independent Component Analysis},
  publisher = {Springer Berlin Heidelberg},
  year = {2007}
}

@BOOK{Comon2010,
  title = {Handbook of blind source seapartion},
  publisher = {Academic Press},
  year = {2010},
  editor = {P. Comon and C. Jutten}
}

@CONFERENCE{Duarte2011,
  author = {Duarte, L. T. and Suyama, R. and Attux, R. and Romano, J. M. T. and
	Jutten, C.},
  title = {Blind extraction of sparse components based on $l_0$-norm minimization},
  booktitle = {Proceedings of the 2011 SSP},
  year = {2011},
  pages = {617-620},
  doi = {10.1109/SSP.2011.5967775},
  issn = {pending}
}

@BOOK{Golub1996,
  title = {Matrix Computations},
  publisher = {John Hopkins},
  year = {1996},
  author = {Golub, Gene H. and Van Loan, Charles F.},
  edition = {3rd}
}

@CONFERENCE{Gribonval2006,
  author = {R. Gribonval and S. Lesage},
  title = {A survey of sparse component analysis for blind source separation:
	principles, perspectives, and new challenges},
  booktitle = {Proceedings of the 2006 ESANN},
  year = {2006},
  pages = {323-330},
  address = {Bruges, Belgium}
}

@BOOK{Hyvarinen2001,
  title = {Independent component analysis},
  publisher = {John Wiley \& Sons},
  year = {2001},
  author = {A. Hyvarinen and J. Karhunen and E. Oja}
}

@InProceedings{Nose-Filho2017,
author="Nose-Filho, K.
and Duarte, L. T.
and Romano, J. M. T.",
editor="Tichavsk{\'y}, Petr
and Babaie-Zadeh, Massoud
and Michel, Olivier J.J.
and Thirion-Moreau, Nad{\`e}ge",
title="On Disjoint Component Analysis",
booktitle="Latent Variable Analysis and Signal Separation",
year="2017",
publisher="Springer International Publishing",
address="Cham",
pages="519--528",
isbn="978-3-319-53547-0"
}

\appendix{}

\section{Proof}
\label{sec:proof}




Relying on the observation that the linear mixing process (invertible) of $N$ bounded sources, with finite real valued amplitudes in the interval $[-A, A]$ will lead to $N$ mixtures with $\ell_\infty$-norm equal or bigger than $A$, we show, in  \textbf{Theorem 1}, that the equalization of the $\ell_\infty$-norm of the estimates recovers the sources.

\textbf{Theorem 1:} Let $\mathbf{s}(n) = [s_1(n) \; s_2(n) \; \cdots \; s_N(n)]^T$ be a set of $N$ bounded sources, as stated in the \ref{sec:intro} and also considering the Extreme Points Condition. Then, any linear combination $$y(n) = \sum_{i=1}^{N}g_i s_i(n),$$ with $||\mathbf{g}||_2 = 1$, will have its infinity norm equalized, $||\mathbf{y}||_\infty = A$, if, and only if, $\mathbf{g} = \mathbf{e}_j$, where $\mathbf{e}_j$ denotes the $j$-th canonical vector, for some $j=1,\cdots, N$. 

\begin{proof}

First, let us start by computing the $\ell_\infty$-norm of $\mathbf{y} = [y(0) \; y(1) \; \cdots \; y(T-1)]^{T}$:

\begin{equation}
    ||\mathbf{y}||_\infty = \underset{n}{\text{max}} \left| \sum_{i=1}^{N}g_i s_i(n) \right|.
\end{equation}

Since the absolute value of a sum is upper bounded by the sum of the absolute values of its terms, and $||\mathbf{s}_i||_\infty = A$, for $i=1, \; \cdots, \;N$, we have \cite{Johnson2012}

\begin{equation}
    \underset{n}{\text{max}} \left| \sum_{i=1}^{N}g_i s_i(n) \right| \leq  \underset{n}{\text{max}} \sum_{i=1}^{N}|g_i| |s_i(n)| \leq A \sum_{i=1}^{N}|g_i|.
    \label{ineq:Ling_ineq}
\end{equation}

From the Extreme Points Condition, the equality in (\ref{ineq:Ling_ineq}) holds
\begin{equation}
    \underset{n}{\text{max}} \left| \sum_{i=1}^{N}g_i s_i(n) \right| = A \sum_{i=1}^{N}|g_i|,
    \label{eq:equality}
\end{equation}
and, consequently,
\begin{equation}
    ||\mathbf{y}||_\infty = A \sum_{i=1}^{N}|g_i|,
\end{equation}
where $\sum_{i=1}^{N}|g_i|$ is the $\ell_1$-norm of $\mathbf{g}$. Considering $||\mathbf{g}||_2 = 1$, from the $\ell_p$ norms inequalities \cite{Foucart2013}, we have
\begin{equation}
    ||\mathbf{g}||_1 \geq ||\mathbf{g}||_2=1,
    \label{eq:norm_ineq_proof}
\end{equation}
and we can find a lower bound for $||\mathbf{y}||_\infty$ 
\begin{equation}
    ||\mathbf{y}||_\infty \geq A,
\end{equation}
where equality holds if, and only if, $||\mathbf{g}||_1 = ||\mathbf{g}||_2$, which means $\mathbf{g} = \mathbf{e}_j$, for some $j=1,\cdots, N$. 
\end{proof}

\textbf{Theorem 1} presents a very interesting result, stating that any linear mixture of bounded sources will have its infinity norm equalized if, and only if, it corresponds to one of the sources. It is important to observe that the Extreme Points Condition is only a sufficient one, as we will also verify in our simulations.

To better understand \textbf{Theorem 1} and the Extreme Points Condition, let us consider the following example.

\par\noindent\rule{\textwidth}{0.4pt}

\textbf{Example:} Let us consider the case of three sources. In such a case

$$y(n) = g_1s_1(n) + g_2s_2(n) + g_3s_3(n)$$
Let us also consider a specific time instant, $n_0$, such that $s_1(n_0) = s_2(n_0) = s_3(n_0) = A$ (from the Extreme Points Condition, such a time instant exists) and another time instant, $n_1$, where $s_1(n_1) = s_2(n_1) = s_3(n_1) = -A$ (the existence of such a time instant follows the same argument). Then, we have

$$y(n_0) = A(g_1 + g_2 +g_3)$$
$$y(n_1) = -A(g_1 + g_2 +g_3)$$

To equalize the $\ell_\infty$-norm of $y(n)$, it will be necessary and sufficient to have

$$g_1 + g_2 +g_3 = 1$$

Since we already have a constraint on the $\ell_2$-norm of $g$, we have the following problem to solve:

$$g_1^{2} + g_{2}^{2} + g_{3}^{2} = 1$$
$$g_1 + g_2 +g_3 = 1$$

We can verify that the canonical vectors $(1,0,0), (0, 1, 0)$ and  $(0,0,1)$ are solutions to these equations, but also the vector, $(\frac{-1}{3}, \frac{2}{3}, \frac{2}{3})$ is a solution.

However, it is important to note that the condition

$$g_1 + g_2 +g_3 = 1$$
came from the time instants $n_0$ and $n_1$, where the 3 sources assumed the values $(A, A, A)$ and $(-A, -A, -A)$. These two conditions are not sufficient to guarantee the Extreme Points Condition. With such conditions, we must consider the time instants where the sources assume the values $(A, A, A), (A, A, -A), (A, -A, A), (A, -A, -A), (-A, A, A), (-A, A, -A)$, $(-A, -A, A), (-A, -A, -A)$.

For example, let us consider a time instant $n_2$ where $s_1(n_2) = -A, s_2(n_2) = s_3(n_2) = A$, which leads to

$$y(n_2) = -A.g_1 + A.g_2 + A.g_3.$$

Using the former solution, $(\frac{-1}{3}, \frac{2}{3}, \frac{2}{3})$ , we have

$$y(n_2) = \frac{A}{3} + 2\frac{A}{3} + 2\frac{A}{3} = 5\frac{A}{3} > A$$
    and the $\ell_\infty$-norm of $y(n)$ is not equal to $A$, which implies that $(\frac{-1}{3}, \frac{2}{3}, \frac{2}{3})$ is no longer a solution.

Hence, the Extreme Points Condition adds constraints to the vector $\mathbf{g}$ that exclude the possibility of spurious solutions, \textit{i.e.}, solutions that equalize the $\ell_\infty$-norm but do not correspond to the canonical vectors. 

\par\noindent\rule{\textwidth}{0.4pt}

We can extend the result obtained for the extraction of one source to the separation of multiple ones. To do so, the mixing matrix must be an orthogonal one, which implies that the sum of the $\ell_\infty$-norms of the mixtures will always be greater than $NA$:
\begin{equation}
    \textrm{J}_{\infty}(\mathbf{W}) = \sum_{i=1}^{N}||\mathbf{y}_i||_\infty = \sum_{i=1}^{N}||{\mathbf{w}_i^{T}}\mathbf{X}||_\infty \geq NA.
    \label{eq:cost_function_J}
\end{equation}
where $\mathbf{w}_i^{T}$ denotes the $i$-th row of $\mathbf{W}$. The equality holds if, and only if, there is exactly one unitary element in each row and column of the global matrix $\mathbf{G}$, \textit{i.e.}, when the sources are separated.

\end{document}